# Decoding Emotional Trajectories: A Temporal-Semantic Network Approach for Latent Depression Assessment in Social Media

Junwei Kuang, Ph.D.
Beijing Institute of Technology
Beijing, China, 100081
Email: kuangjw@bit.edu.cn

Jiaheng Xie, Ph.D.
University of Delaware
217 Purnell Hall
Newark, DE, USA 19716
Email: jxie@udel.edu

Zhijun Yan, Ph.D.
Beijing Institute of Technology
Beijing, China, 100081
Email: yanzhijun@bit.edu.cn

Please send comments to Junwei Kuang at kuangjw@bit.edu.cn

# Decoding Emotional Trajectories: A Temporal-Semantic Network Approach for Latent Depression Assessment in Social Media

## Abstract

The early identification and intervention of latent depression are of significant societal importance for mental health governance. While current automated detection methods based on social media have shown progress, their decision-making processes often lack a clinically interpretable framework, particularly in capturing the duration and dynamic evolution of depressive symptoms. To address this, this study introduces a semantic parsing network integrated with multi-scale temporal prototype learning. The model detects depressive states by capturing temporal patterns and semantic prototypes in users' emotional expression, providing a duration-aware interpretation of underlying symptoms. Validated on a large-scale social media dataset, the model outperforms existing state-of-the-art methods. Analytical results indicate that the model can identify emotional expression patterns not systematically documented in traditional survey-based approaches, such as sustained narratives expressing admiration for an "alternative life." Further user evaluation demonstrates the model's superior interpretability compared to baseline methods. This research contributes a structurally transparent, clinically aligned framework for depression detection in social media to the information systems literature. In practice, the model can generate dynamic emotional profiles for social platform users, assisting in the targeted allocation of mental health support resources.

# Decoding Emotional Trajectories: A Temporal-Semantic Network Approach for Latent Depression Assessment in Social Media

## INTRODUCTION

Mental health is a critical and integral component of overall health, and depression is one of the most prevalent mental disorders (WHO, 2017). Approximately 280 million people suffer from depression worldwide (Murray, 2022), and global depression cases have soared by 28% since the outbreak of the COVID-19 pandemic (Santomauro et al., 2021). Such a sizable and rapidly growing depression population has brought significant societal and financial consequences. More than one million people worldwide commit suicide due to depression annually, on par with the number of deaths from cancer (WHO, 2017). The economic toll linked to depression increased by 37.9% from $236.6 billion to $326.2 billion during 2010-2018 in the US (Greenberg et al., 2021) and is projected to be the world's leading economic burden by 2030 (WHO, 2017). While many effective depression treatments exist, more than 70% of patients do not seek treatments due to stigmatization around depression (Shen et al., 2017). To mitigate this societal issue and avoid preventable ramifications, depression detection is the key (Picardi et al., 2016).

Survey-based methods hold the gold standard for clinical depression detection, which examines the frequency of *depressive symptoms in the past two weeks* according to clinical depression screening methods such as PHQ-9 (Kroenke et al., 2001). Beyond such a small scale of offline detection, social media, representing 58.4% of the world's population (DataReportal, 2022), unleashes the unprecedented potential to expand its reach to the online setting while scaling up patient coverage. Social media allow users to create and share user-generated content (Aichner et al., 2021), which offer an authentic and comprehensive landscape of patients' historical conditions (Salas-Zárate et al., 2022). Moreover, depressed patients are more willing to communicate on social media compared to offline due to the online disinhibition effect (Naslund et al., 2016). To make use of such an indispensable resource, many scholars develop depression detection models on social media for early intervention (Chau et al., 2020; Liu et al., 2022). Although achieving satisfying performance, most of these studies rely on black-box methods. Abundant evidence shows that lack of interpretability results in limited applicability and potential risk in high-stake scenarios such as healthcare decision-making (Chiong et al., 2021b; Zogan et al., 2022). For example, Zech et al. (2018) report training a medical disease prediction model based on x-rays, but the model keyed on the meta-tagged word "portable" which is reflective of where the samples came from instead of a valid signal of disease. To overcome the non-interpretable dilemma, a few depression detection studies attempt to explain why users are classified as depressed based on the importance score or attention weights of interpretable inputs such as words in a post (Cheng & Chen, 2022). One recent IS study closest to our task is Chau et al. (2020), who detect depression in social media by combining rule-based classification. For example, if a sentence contains negative emotion words and does not contain negation words, then the post indicates depression. These rules are based on linguistic features, which depart from the clinical diagnosis criterion that is symptom-based. Unable to reveal those depressive symptoms, the existing interpretable models receive compromised trust from end users, fall short of lending personalized support to patients, and are impractical in deriving clinical-based interventions. To tackle their limitations, there has recently been a rising interest in utilizing symptoms for interpreting depression detection. Pioneering studies have shown the potential benefits of improving accuracy, generalizability, and interpretability (Nguyen et al., 2022; Zhang et al., 2022b). Therefore, our research objective is to develop an ***interpretable depression detection model in social media based on symptom-based depression diagnostic criteria***.

The symptom-based interpretable methods for depression detection can be categorized into dictionary-based, similarity-based, and classification-based. The core of these methods is to identify depressive symptoms from user-generated posts on social media. Dictionary-based methods first collect symptom-related words as a dictionary and then identify symptoms based on the number of these words in posts (Shen et al., 2017). Similarity-based methods examine depressive symptoms based on the embedding similarity of posts and symptom templates, which are hand-crafted descriptions such as “I am disappointed in myself.” Classification-based methods treat pre-defined symptoms as the prediction outcome (Zhang et al., 2022b). However, these methods still face three limitations. First, prior methods only identify pre-defined symptoms according to the literature. However, depressive symptoms may evolve over time. For example, the Diagnostic and Statistical Manual of mental disorders (DSM) took 12 years to update from DSM-IV to DSM-V to accommodate symptom evolution (Shen et al., 2017). Second, previous methods rely on domain-specific knowledge such as dictionaries and templates to examine symptoms, which require significant labor costs and suffer from poor generalizability. Third, extant methods focus on *what symptoms* users present while neglecting *how long* these symptoms last, which is equally critical for clinical depression diagnosis (Kroenke et al., 2001). Frequent depressive episodes or the long duration of a single episode are significant signals of depression (Herrman et al., 2022). Fortunately, user-generated posts on social media can reveal such “how long” aspects of depressive symptoms. As shown in Figure 1, the user reported the disturbed sleep symptom numerous times, ranging from Feb 14 to May 14. Certain periods (e.g., May 12 to May 14) show denser symptom mentions than others. This “how long” information can be fruitfully leveraged to guide the method design and improve the predictive power and interpretability of depression detection.

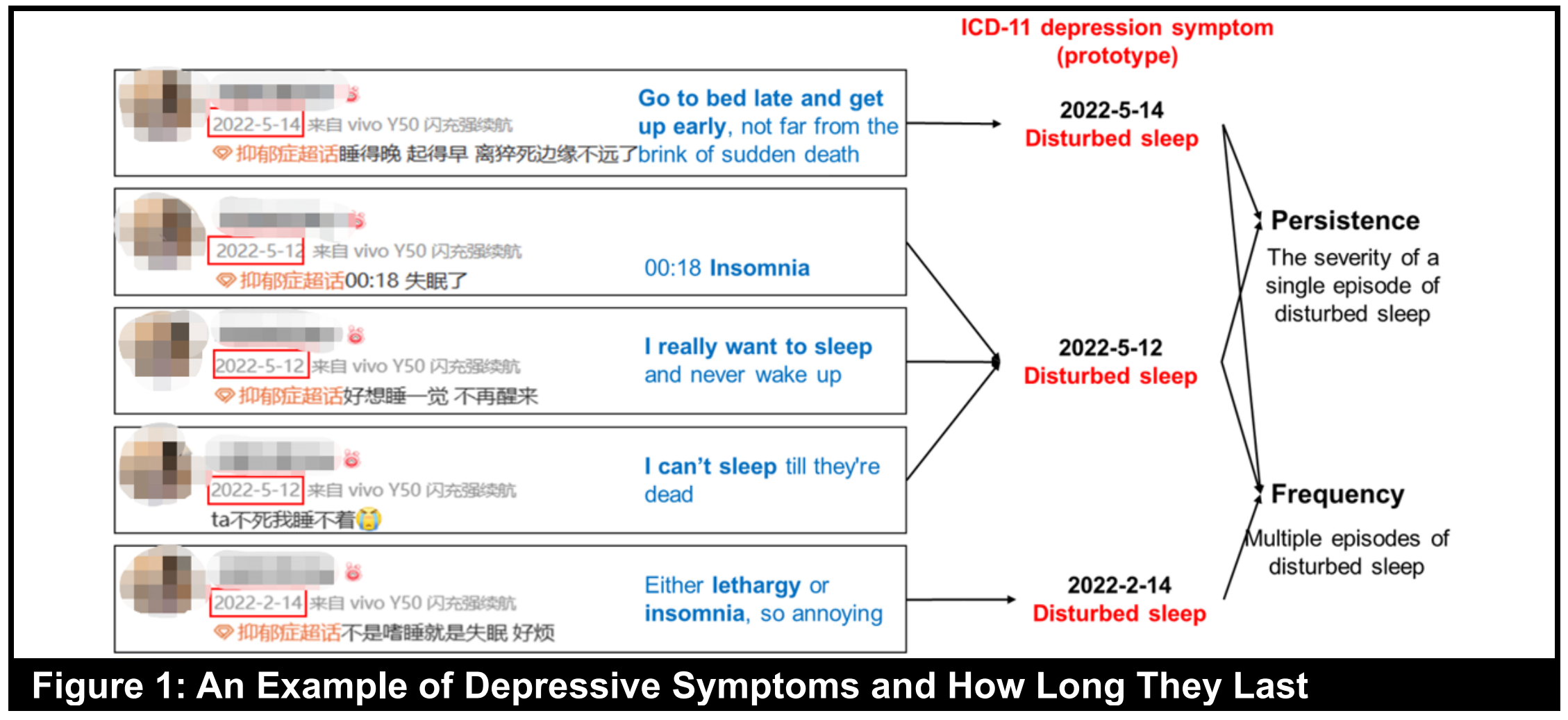

**Figure 1: An Example of Depressive Symptoms and How Long They Last**

The abovementioned limitations motivate us to develop a novel interpretable depression detection method that is capable of discovering depressive symptoms in a data-driven manner while capturing how long these symptoms last. Following the computational design science paradigm and prior IS research on health analytics (Ben-Assuli & Padman, 2020; Lin et al., 2017; Yu et al., 2023), we propose and rigorously evaluate a novel interpretable model, Multi-Scale Temporal Prototype Network (MSTPNet). MSTPNet is built upon an emergent stream of case-based interpretable models, prototype learning (Ming et al., 2019), which interprets the prediction for new inputs by comparing them with a few learned prototypes. In this study, typical posts disclosing depressive symptoms can be recognized as prototypes. These prototypes are automatically learned depending on the predicted outcome without pre-defining them and can be seamlessly adapted for symptom identification at low labor costs. To consider how long the symptoms last, MSTPNet modifies standard prototype learning methods by devising two novel layers: a temporal segmentation layer that eliminates the negative effects of irrelevant and redundant posts on symptom identification to facilitate period-level analysis (i.e., "What symptoms did the user suffer in a period?"), and a multi-scale temporal prototype layer that captures the frequency and persistence of symptoms over different consecutive periods.

This study contributes to healthcare IS research with a novel interpretable depression detection method using social media with clinically relevant interpretation. Our empirical findings uncover new depressive symptoms unnoted in previous medical literature, such as sharing admiration for a different life. This study also contributes to the interpretable machine learning literature with a novel prototype learning method, which can process a sequence of user-generated posts and interprets its decision based on the type and temporal distribution of the prototype. Our proposed method is generalizable for other applications including time-sensitive prototypes, such as sensor-based disease detection. In practice, our proposed method can be implemented in social media platforms to detect depressed patients and interpret their temporal symptoms, which contribute to the personalized intervention in the corresponding symptoms.

# LITERATURE REVIEW

## Depression Detection in Social Media

Social media-based depression detection is broadly classified into post-level detection and user-level detection. The post-level detection aims to predict whether a post contains depression-related emotions (Chiong et al., 2021a). The user-level detection focuses on whether a user is depressed (Malhotra & Jindal, 2022). Our study belongs to the latter category, as it is more clinically relevant and more direct for interventions. The earliest user-level detection method is survey-based, which deploys surveys to social media users (Park et al., 2012). These survey-based methods are labor-intensive, time-consuming, and hard to scale up. To overcome those challenges, the mainstream approach is automated methods that can be categorized into traditional machine learning, black-box deep learning, and interpretable deep learning.

### Traditional Machine Learning Methods in Depression Detection

Traditional machine learning methods refer to machine learning methods such as Support Vector Machine (SVM) other than deep learning (Zhou, 2021). The traditional machine learning-based depression detection model mostly relies on effective input features (Li et al., 2019). The majority of input features are extracted from user-generated posts using Natural Language Processing (NLP) techniques, which are guided by theories of depression (Zulkarnain et al., 2020). For example, due to too much self-awareness and less connection to others, the frequent use of first-person singular words, such as "I," are often used to detect depression (Bucci & Freedman, 1981). Previous studies also have found more frequent use of negative emotional words, and absolutist words among depressed patients (Li et al., 2019). These linguistic features can be captured by Linguistic Inquiry and Word Count (LIWC) and Part of Speech (POS) techniques (Bucur et al., 2021; Pennebaker et al., 1999). In addition to linguistic features, the semantics of the user's posts are also critical for depression detection. Such important semantics generally are extracted by N-gram and Latent Dirichlet Allocation (LDA) models (Zogan et al., 2022). Beyond the above standard textual features, a few studies design domain-specific features, such as antidepressant words count per post (Zulkarnain et al., 2020). In addition to text data, a few studies extract features from other data modalities, such as images, and posting time (Liu et al., 2022). Table 1 summarizes the traditional machine learning-based method.

**Table 1: Traditional Machine Learning Methods in Social Media-based Depression Detection**

| Reference | Dataset | Sample (depression/non) | Input features | Methods |
|---|---|---|---|---|
| Choudhury et al. (2013) | Twitter | 476 (171/305) | Emotion, Depression language, Language style | SVM |
| Tsugawa et al. (2015) | Twitter | 209 (81/128) | Emotions, Linguistic style, Topic, social Network | LDA, SVM |
| Shen et al. (2017) | Twitter | 2804 (1402/1402) | Social Network, Emotional, Topic, Domain-specific | MDL, NB |
| Sadeque et al. (2018) | Reddit | 888 (136/752) | Depression lexicon, Meta features | LibSVM, Ensemble, |
| Chen et al. (2018) | Twitter | 1200 (600/600) | Emotion swings, LIWC | SVM, RF |
| Chau et al. (2020) | Blog | 804 (274/530) | N-gram, Lexicon based, LIWC | SVM, Rule-based, GA |
| Chiong et al. (2021b) | Twitter | 2804 (1402/1402) | N-gram | SVM, DT, NB, KNN, RF |

However, these studies have shown unsatisfactory predictive power, arising from three limitations. First, hand-crafted features are limited by domain knowledge, so it is difficult to capture sufficient input features. Second, depressed people often use informal and implicit words to express their negative emotions, which are not captured by hand-crafted features. Third, traditional machine learning models are not complex enough to capture high-level interactions between features. Deep learning can address the above limitations.

**Black-Box Deep Learning Methods in Depression Detection**

Black-Box deep learning methods refer to deep learning methods without interpretability and have demonstrated significantly higher predictive power in depression detection compared to traditional machine learning methods (Malhotra & Jindal, 2022). These improvements have benefited from the development of embedding techniques and the utilization of various neural network architectures. As depressed patients often use informal words and implicit words to express their emotions, embedding techniques are used to capture accurate and rich semantics (Pérez et al., 2022). Given that depressed patients have a sequence of posts, RNN models, especially Long Short Term Memory (LSTM) models, have become essential components to capture the evolution of patients' moods and symptoms over time (Ghosh & Anwar, 2021). A few studies have also utilized CNNs to process textual data to extract local features, such as depression-related phrases, as well as global features for depression detection (Wang et al., 2022). To enhance the model's predictive power by capturing important clues related to depression, depression detection methods with attention mechanisms such as Hierarchical Attention Network (HAN) have been proposed (Cheng & Chen, 2022). Table 2 summarizes recent black-box deep learning-based depression detection methods in social media. These methods leverage multi-modal data and have been evaluated on different datasets with varying imbalance ratios from 2:1 to 1:6 (depression: non-depression).

Despite their satisfying performance, their lack of interpretability limits their applicability in high-stake decision-making scenarios (Rudin, 2019). The interpretability of the model is essential to increase trust in a model, prevent failures, and justify its usage (Moss et al., 2022). On one hand, without understanding the rationale of a prediction, it is difficult to judge exactly how it will perform in new scenarios. The good performance of black box models in the training set may come from noise, which has poor generalizability. On the other hand, given the complexity and heterogeneity of depression, each patient with depression is unique and needs personalized treatment (Herrman et al., 2022). However, the black box model does not provide insights into the understanding of depressed patients, and inappropriate interventions may deteriorate the condition (Mikal et al., 2016). For instance, if the platform recommends articles about the dangers of suicide to patients who do not have any suicidal thoughts, these articles could potentially trigger suicidal thoughts in patients. Moreover, the General Data Protection Regulation (GDPR) in the European Union has enforced "the right to explanation" for individual-level prediction algorithms.

**Table 2: Black-Box Deep Learning-based Methods in Social Media-based Depression Detection**

| Reference | Dataset | Sample (depression/non) | Input | Methods |
|---|---|---|---|---|
| Orabi et al. (2018) | Twitter | 899 (327/572) | Text | CNN/RNN |
| Chiu et al. (2021) | Instagram | 520 (260/260) | Text, Image, Posting time | LSTM with temporal weighting and day-based aggregations |
| Ghosh and Anwar (2021) | Twitter | 6562 (1402/5160) | Text | LSTM |
| Zhang et al. (2022a) | Reddit | 892 (137/755) | Text | Hierarchical Attention Network |
| Zogan et al. (2022) | Twitter | 4800 (2500/2300) | Text, Image | Hierarchical Attention Network |
| Kour and Gupta (2022) | Twitter | 1681 (941/740) | Text | CNN + Bi-LSTM |
| Wang et al. (2022) | Weibo | 32570 (10325/22245) | Text, Image, Posting time | Multitask learning |
| Naseem et al. (2022) | Reddit | 190 (125/65) | Text | Bi-LSTM with an attention mechanism |
| Cheng and Chen (2022) | Instagram | 1054 (526/528) | Text, Images, Posting time | Time Aware LSTM with Attention |

### Interpretable Deep Learning Methods in Depression Detection

There are two main definitions of interpretability in social media analytics (Dhurandhar et al., 2017; Lee & Yoon, 2019). One definition is the degree to which a user can understand and trust the cause of a decision (Molnar, 2020). Another definition suggests “AI is interpretable to the extent that the produced interpretation can maximize a user’s target performance” (Dhurandhar et al., 2017). Interpretable deep learning methods refer to deep learning methods that provide a certain explanation (Li et al., 2022). Prevailing interpretable deep learning methods applied to social media-based depression detection are generally classified into the following three types.

Approximation-based interpretable deep learning methods in depression detection provide explanations by revealing the importance score and direction of interpretable input (e.g., words in a post) linked to the final output (Li et al., 2022). Since deep learning models are generally not inherently interpretable, these explanations are generated by post hoc explanation methods. These post hoc methods interpret individual model predictions by learning a simple interpretable model (e.g., linear regression) locally approximating the original model around a given prediction (Ras et al., 2022). The popular post hoc explanation methods include Shapley additive explanations (SHAP) and Local Interpretable Model-agnostic Explanations (LIME) (Lundberg & Lee, 2017; Ribeiro et al., 2016). These methods assign an importance score to input words or posts and visualize them by highlighting key clues for depression detection. End users compare these explanations with their intuitions or knowledge of depression detection to determine whether they trust the prediction or the model. For example, Adarsh et al. (2023) provide a list of specific depression-related words (e.g., sad, lonely, and stressed) using LIME as an explanation. However, these methods are vulnerable because the explanation is from the explanation model instead of the actual knowledge from the data (Slack et al., 2020). Zeiler and Fergus (2014) report that the random perturbation that LIME results in unstable interpretations. To overcome this issue, experts have been advocating intrinsic explanations of model *per se* (Rudin, 2019).

Attention-based interpretable deep learning methods are inherently interpretable, which show how input elements influence the model’s later processing and final decisions based on attention mechanisms (Hu, 2020). Moreover, deep learning with attention mechanisms can simultaneously preserve or even improve the predictive power of depression detection and provide explanations based on attention weights (Ras et al., 2022). In social media-based depression detection, attention mechanisms enable the model to attend to important words within a post and important posts (Cheng & Chen, 2022). However, the method faces two limitations. First, only highlighting which inputs are important without sufficient clarity on how they correlate with the prediction task confuses end-users (Payrovnaziri et al., 2020). Moreover, the attention weights are frequently so diffused that a multitude of inputs are concurrently highlighted. Not only would this obfuscate the identities of the truly important inputs but would also result in information overload for the end-users. Second, the interpretable input-based explanations are inconsistent with the clinical depression diagnosis criterion that is based on depressive symptoms. Such misalignment results in compromised trust from end users and fall short of facilitating the cooperation of human experts and AI models. Given the complexity of depression detection involving a sequence of posts, it is easier to understand and trust by extracting depressive symptoms and then providing symptom-based explanations, rather than explanations based on the raw input elements (Lee et al., 2018).

Symptom-based interpretable deep learning methods in depression detection aim to provide symptom-based explanations by identifying depressive symptoms from user-generated posts (Zhang et al., 2022b). There has recently been a rising interest in symptom-based interpretable methods, which can be categorized according to the symptom identification methods into dictionary-based, similarity-based, and classification-based. The dictionary-based method first constructs dictionaries containing specific keywords for each depressive symptom (Mowery et al., 2017). A user-generated post is then associated with one or more depressive symptoms based on the count of symptom-related words (Shen et al., 2017). The similarity-based method employs depression templates derived from established depression scales to identify symptoms based on the embedding similarity of posts and templates (Zhang et al., 2022a). These templates include direct expressions of depressive moods and depression treatments, as well as theory-grounded indirect symptoms such as guilty feelings and pessimism (Ahmed et al., 2022). Given a post, the most similar template serves as its diagnostic basis, and the similarity is considered an indicator of the post's risk level. The classification-based method builds a classifier for each depressive symptom or a multi-label classifier to predict the relevance of each post-symptom pair (Zhang et al., 2022b). Table 3 summarizes and contrasts recent interpretable deep learning-based methods and our study in social media-based depression detection.

**Table 3: Interpretable Deep Learning Methods in Social Media-based Depression Detection**

| Reference | Type | Method | Usage | Explanations |
|---|---|---|---|---|
| Adarsh et al. (2023) | Approximation | LIME | Post-hoc | Important raw inputs |
| Bucur et al. (2021) | Approximation | SHAP | Post-hoc | Important raw inputs |
| Cheng and Chen (2022) | Attention | Attention | Intrinsic | Important raw inputs |
| Zogan et al. (2022) | Attention | HAN | Intrinsic | Important raw inputs |
| Shen et al. (2017) | Symptom | Dictionary-based | Post-hoc | Predicted symptoms |
| Zhang et al. (2022b) | Symptom | Classification-based | Post-hoc | Predicted symptoms |
| Zhang et al. (2022a) | Symptom | Similarity-based | Post-hoc | Predicted symptoms |
| **Our study** | **Symptom** | **Similarity-based** | **Intrinsic** | **More predicted symptoms, and how long they last** |

However, current symptom-based methods still face two limitations. First, they generally require high labor costs and only identify pre-defined symptoms from the clinical depression diagnosis criterion, neglecting new symptoms unnoted in offline depression screening questionnaires in the online setting. Second, symptom-based interpretable methods focus only on the type of depressive symptoms users suffer, neglecting how long these symptoms last, which is equally critical for a clinical depression diagnosis. These limitations motivate us to develop a novel interpretable depression detection method that is capable of discovering depressive symptoms in a data-driven manner while capturing how long these symptoms last.

## Prototype Learning Methods

To understand depressive symptoms as the basis for the interpretation of depression detection, we resort to an emergent interpretable model paradigm that is closely related to our task: prototype learning. Prototype learning methods learn prototypes that have clear semantic meanings, and intrinsic explanations are generated based on the comparison between input and each prototype (Nauta et al., 2021). Prototype learning was originally proposed to interpret image recognition. Chen et al. (2019) propose ProtoPNet (Figure 2), which explains the contribution of prototypical parts of the predicted image by comparing the learned prototypes.

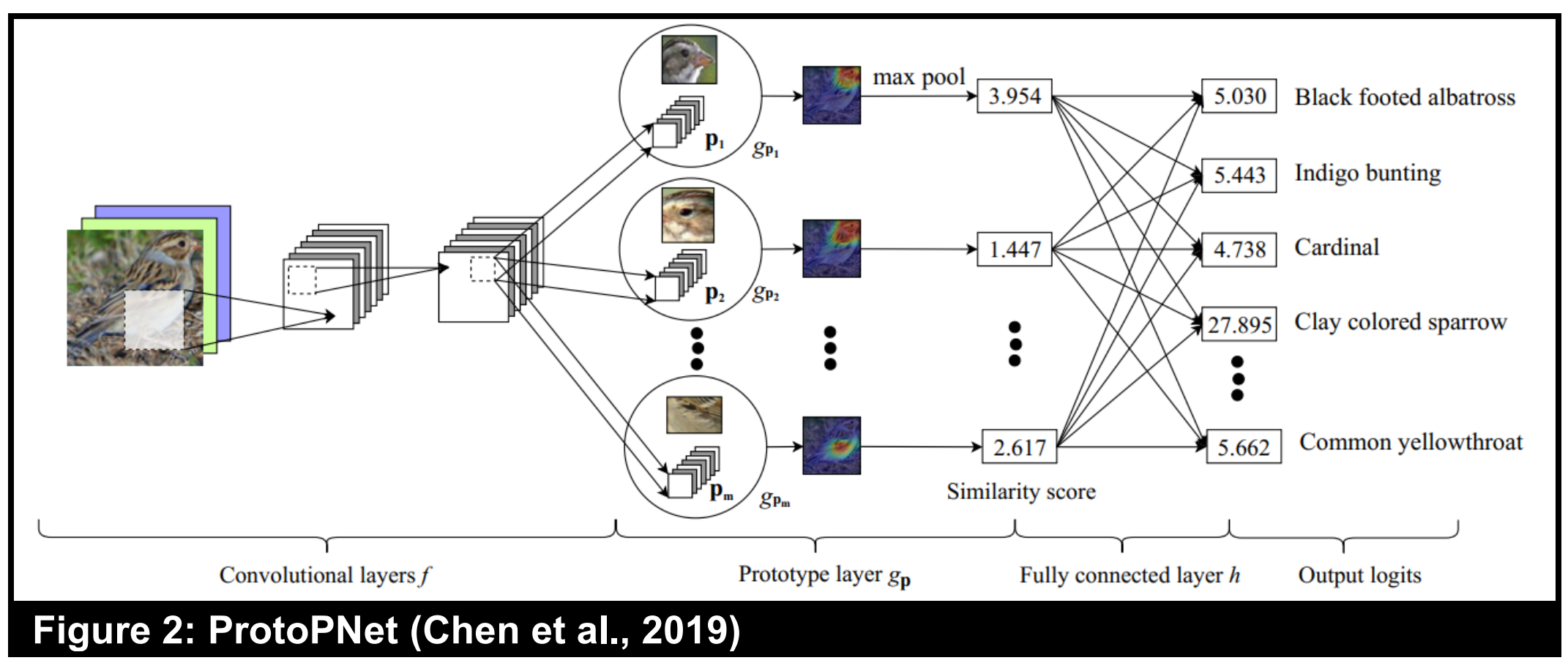

**Figure 2: ProtoPNet (Chen et al., 2019)**

This model introduces a prototype layer with $k$ prototypes, where each prototype is the most salient and typical representation, such as the head of a clay-colored sparrow. ProtoPNet embeds each prototype $j$ as a vector $p_j$ in the latent space and defines *prototype similarity* $s_j$ to measure how strongly prototype $j$ exists in the input bird picture by comparing $p_j$ and the feature maps of the picture. The model subsequently classifies the input bird picture based on the weighted sum of the prototype similarities computed between this picture and each prototype. Prototype learning has been extended to interpret other machine learning tasks such as text classification (e.g., sentiment analysis), where typical positive and negative posts are learned as prototypes (Ming et al., 2019), video classification (Trinh et al., 2021), time series classification (Zhang et al., 2020). Prototype learning has also been enhanced by combining attention mechanisms (Zhang et al., 2020), hierarchies (Hase et al., 2019), and federated learning (Tan et al., 2022). Table 4 summarizes and contrasts major prototype learning methods with our method.

**Table 4: Existing Prototype Learning Methods vs. Our Method**

| Reference | Method | Novelty | Input | TD* |
|---|---|---|---|---|
| Chen et al. (2019) | ProtoPNet | Prototype for image classification | An image | No |
| Hase et al. (2019) | HPNet | Hierarchical prototype | An image | No |
| Ming et al. (2019) | ProSeNet | Prototype for text classification | A piece of text | No |
| Gee et al. (2019) | Autoencoder and prototype | Prototype for time series classification | A time series of ECG | No |
| Hong et al. (2020) | ProtoryNet | Prototype trajectories | A piece of text | No |
| Zhang et al. (2020) | TapNet | Attentional prototype | A time series of ECG | No |
| Rymarczyk et al. (2021) | ProtoPShare | Prototype parts share | An image | No |
| Nauta et al. (2021) | ProtoTree | Prototype and decision tree | An image | No |
| Trinh et al. (2021) | DPNet | Dynamic prototype | A clip of a video | No |
| Tan et al. (2022) | FedProto | Federated learning prototype | An image or a piece of text | No |
| Deng et al. (2022) | K-HPN | Pairwise prototype | A piece of text | No |
| **Our study** | **MSTPNet** | **Multi-scale temporal prototypes** | **A sequence of text documents** | **Yes** |

* TD stands for "Temporal Distribution," indicating whether a model considers the temporal distribution of the prototype, which includes frequency and persistence of appearance at the period level.

Prototype learning is a suitable method to extract depressive symptoms in our study. Patients may disclose a variety of depressive symptoms in their posts. Typical posts disclosing depressive symptoms can be recognized as prototypes. By calculating how similar a user's posts are to these prototypes, this user's depressive symptoms can be inferred, which serves as a natural interpretation mechanism. However, these prototype learning methods are still limited in terms of capturing the temporal distribution of depressive symptoms. The majority of prototype learning methods focus on static subjects, such as an image and a piece of text. When applied to our study, these methods only consider whether depressive symptoms appear, neglecting how long each symptom last. While a few prototype learning methods process dynamic subjects such as video and ECG signals, these methods focus on directly identifying complex prototypes with temporal properties, rather than analyzing the temporal distribution of prototypes after identifying them. Our method aims to incorporate the temporal distribution of symptoms into the prototype learning method to effectively capture how long depressive symptoms last to improve the predictive power and interpretability of depression detection in social media.

## Temporal Distribution of Symptoms

Temporal distribution describes the distribution of events over periods, often used to analyze and predict trends. Social media analytics has unleashed the unprecedented potential for population-level monitoring and user-level prediction, where temporal distribution plays an important role.

Population-level monitoring aims to infer the evolution of important events such as emotion and disease in certain populations by analyzing the content and time of users' posts on social media. For example, Gruebner et al. (2018) use Twitter to extract negative emotions indicating discomfort in New York City before, during, and after Superstorm Sandy in 2012. Ye et al. (2016) report the temporal development and evolution of dengue fever in China using messages from Weibo by constructing a list of keywords related to dengue fever to analyze how frequently these words appear in Weibo messages. User-level prediction involves classifying social media users. For example, Kokkodis et al. (2020) use the HMM model to predict whether a user is a community contributor or lurker in the future based on the user's past community engagement activities such as posting and replying. Our study belongs to the user-level prediction.

The temporal distribution of symptoms is indispensable to diagnostic criteria for clinical depression (Kroenke et al., 2001). While healthy people could suffer from temporary depressive symptoms in adversity, depressed patients experience episodes more frequently and each episode lasts longer (Herrman et al., 2022). Despite its value, the temporal distribution of depressive symptoms has been largely overlooked in social media-based depression detection. Moreover, current user-level detection methods are still limited in our study. First, the predicted subjects of these methods are online, such as bot users (Costa et al., 2015) and community contributors (Kokkodis et al., 2020), where user-generated posts as predictors are relevant and complete. However, users' posts are only proxies for offline depression conditions and may be irrelevant or incomplete for depression detection. For example, people with depression may not share enough posts during a depressive episode. It is challenging to learn the real health status from a limited number of online posts. Second, these methods mostly focus on the measurement of frequency, such as the percentage of posts at late night (Wang et al., 2022), neglecting the measurement of persistence, which indicates the degree to which something (e.g., depressive symptoms) happens continuously. To overcome these limitations, our method aims to more accurately and comprehensively capture the temporal distribution of depressive symptoms.

## Key Novelties of Our Study

The above literature review reveals multiple research gaps and opportunities for method innovation. On one hand, prior symptom-based interpretable methods in depression detection mostly require high labor costs and only identify pre-defined depressive symptoms. To address this limitation, we build upon prototype learning to explore and identify depressive symptoms from user-generated posts in a data-driven manner. On the other hand, prevailing prototype learning methods focus on whether the prototype appears in the input elements but generally overlook how long these prototypes last, which is essential to improve the predictive power and interpretability of social media-based depression detection. These limitations motivate us to develop a novel prototype learning method, which can show what symptoms the user presents and how long these symptoms last from user-generated posts. However, the "how long" of depressive symptoms is non-trivial to measure. First, our method needs to avoid the negative effect of distracting posts, including depression-unrelated posts and depression-related but redundant posts, to get an accurate and comprehensive landscape of patients' symptoms at different periods. Second, we need effectively measures the persistence of depressive symptoms.

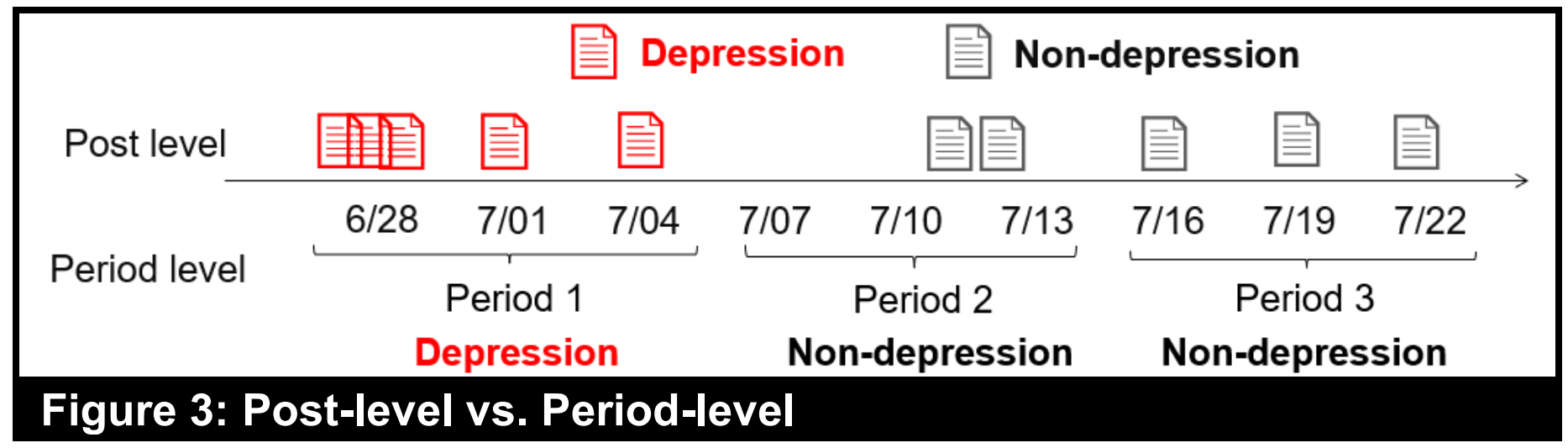

**Figure 3: Post-level vs. Period-level**

To capture the temporal distribution of depressive symptoms, the prerequisite is to know the onset of symptoms at different periods. In response, we propose the period-level analysis for depression detection, which is different from current user-level and post-level depression detection. As shown in Figure 3, while the user has five depression posts and five non-depression posts, the figure shows that the user has depression posts in one out of three consecutive periods from a period-level perspective. The period-level analysis avoids overestimating the effect of redundant posts on depression detection. To divide similar posts into a period, we propose a temporal segmentation layer that considers semantic similarity and time intervals between posts.

Based on the presence of symptoms at different periods, we explicitly capture important temporal measurements such as frequency and persistence, which are interpretable for depression detection. For frequency, there is a categorical difference between one and multiple episodes of depressive symptoms, which distinguishes temporary mood swings from major depressive disorder (WHO, 2019). Depressive symptoms appearing in multiple continuous periods (i.e., persistence) are also a significant sign of depression, which existing methods fall short of. To fill the gap, our method measures the persistence of depressive symptoms by using a sliding window-like approach to traverse the onset of depressive symptoms over continuous periods with different lengths (i.e., multi-scale). The different scales can capture temporal measurements of symptoms at different granularity levels. Based on these observations and designs, our method innovatively detects and interprets depressive symptoms as well as how long they last.

## PROBLEM FORMULATION

For a given user $u \in U$ in social media, let $y^{(u)}$ be this user's depression status, where $y^{(u)} = 1$ denotes depression, and $y^{(u)} = 0$ represent non-depression. For each user $u$, we also observe this user's $N_u$ posts denoted by $X^{(u)} = \left(X_1^{(u)}, X_2^{(u)}, \dots, X_{N_u}^{(u)}\right)$ that were published at time $T^{(u)} = \left(t_1^{(u)}, t_2^{(u)}, \dots, t_{N_u}^{(u)}\right)$. We observe the dataset $D = \left\{\left(X^{(u)}, T^{(u)}, y^{(u)}\right) | u = 1,2, \dots, |U|\right\}$, and $|U|$ denotes the number of users. The objective of the social-media-based interpretable depression detection problem is to learn a predictive model from $D$ that can predict depression status of each user $u$ and interpret the prediction. Table 5 summarizes the important notations.

**Table 5: Important Notations**

| Notation | Description | Notation | Description |
|---|---|---|---|
| $X_i$ | The $i$th post of a focal user | $t_i$ | The time when the post $X_i$ is published |
| $y$ | Depression status of focal user, 1 denotes depression, else non-depression | $p_k$ | The embedding vector of the $k$-th symptom prototype |
| $C_m$ | The $m$th cluster of a focal user | $X_{m,l}$ | The $l$th post of the $m$th cluster |
| $H_{m,l}$ | The embedding vector of the post $X_{m,n}$ | $s_{m,k}$ | The existence strength of symptom prototype $k$ in clusters $C_m$ |
| $s_{m,l,k}$ | The existence strength of symptom prototype $k$ in post $X_{m,l}$ | $g_{j,k}$ | The existence strength of symptom prototype $k$ in scale $j$ |
| $w_j$ | The value of the $j$-th scale | | |

To solve this problem and provide symptom-based explanations, it is critical to learn what symptoms the user presents and how long these symptoms last from user-generated posts. To learn this critical information, we need to tackle two major technique challenges. First, unlike real-time sequence data such as ECG and sensor signals, the user-generated posts are discrete and irregularly distributed, which makes it difficult to measure how long users' depressive symptoms last. Although previous studies use the number of depressive symptom-related posts to measure the "how long" aspect, this measurement may be overestimated because users may generate too many posts during a single depressive episode (Shen et al., 2017). Second, it is difficult to capture the temporal distribution of depressive symptoms from user-generated posts using existing prototype learning methods. The input of most prototype learning methods is a static object such as an image or a piece of text, which fall short of capturing the temporal distribution of prototypes. Although a few variants of prototype learning methods consider temporal distribution, they recognize the temporal distribution as a new prototype, such as a bradycardia prototype represented by a segment of ECG signal fluctuation. If applying these methods to our study, a segment of adjacent posts will be used to learn these complex prototypes such as short-term loss of interest symptoms. Such a large number of potential prototypes make it infeasible to learn. Therefore, it is necessary to propose a new method to effectively capture the "how long" aspects of depressive symptoms and address the above challenges.

## MULTI-SCALE TEMPORAL PROTOTYPE NETWORK

We propose a novel interpretable deep learning method, Multi-Scale Temporal Prototype Network (MSTPNet), to detect depression in social media and provide symptom-based interpretation. Figure 4 shows the architecture of MSTPNet, which features four building blocks. The feature learning layer aims to represent each post as an embedding vector with a fixed length and rich semantic meaning. Different from analyzing each post independently, our proposed temporal segmentation layer assigns posts into different periods based on the semantic similarity and time interval between posts, which facilitate period-level analysis to support the following measurement of the "how long" aspect. Instead of learning complex dynamic prototypes (e.g., "long-term disturbed sleep") directly, our proposed multi-scale temporal prototype layer breaks the task down into two parts. We first infer depressive symptoms (e.g., "disturbed sleep") in each period by comparing posts with learned prototypes, and then explicitly measure the frequency (e.g., the proportion of periods where disturbed sleep appears) and persistence (e.g., the number of continuous periods where disturbed sleep all appears) of each symptom. Based on the above interpretable temporal measurement of each symptom, the classification layer classifies a user into depression or non-depression categories. The following subsections describe each block in order, Lastly, we illustrate the reasoning process of our network's interpretation.

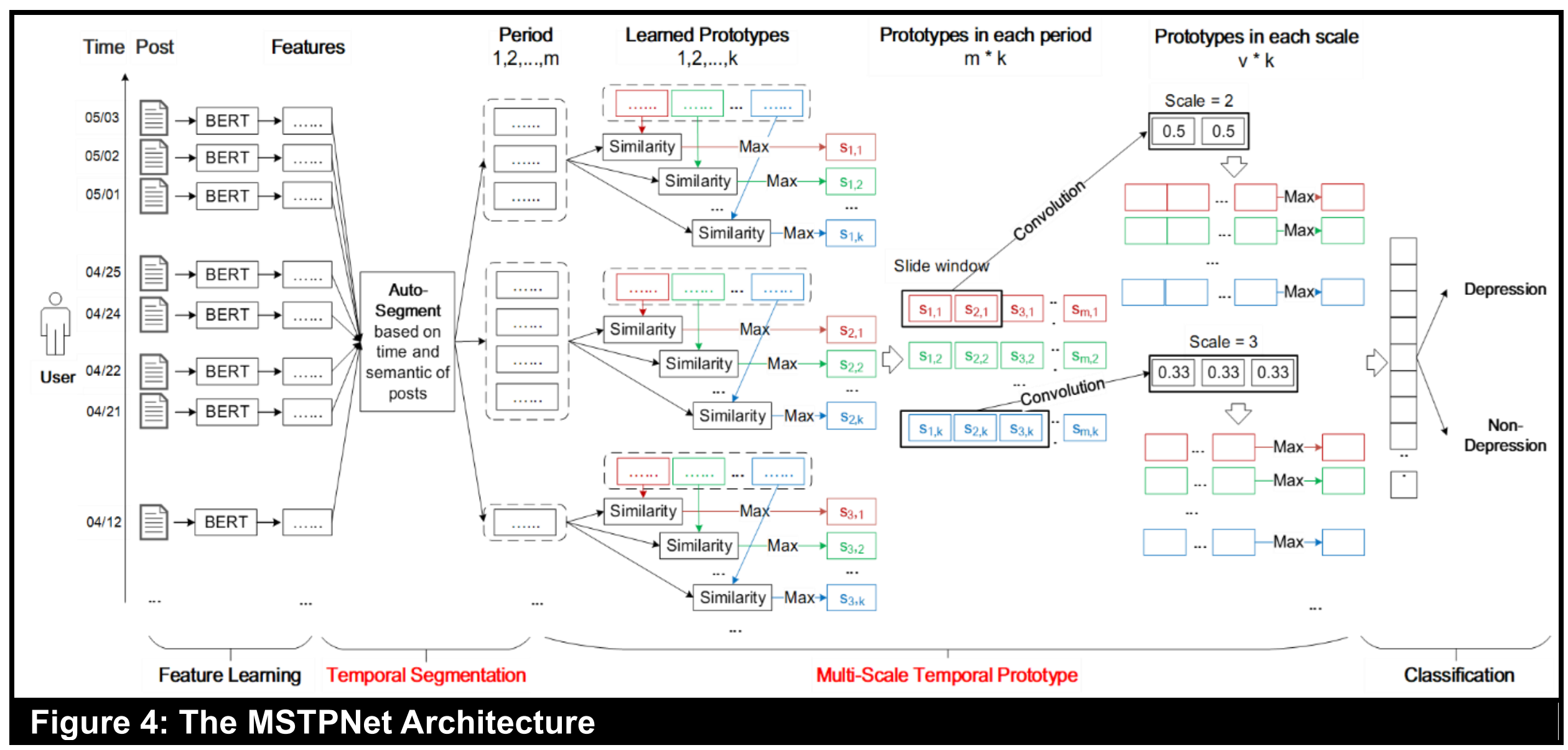

Figure 4: The MSTPNet Architecture

## Feature Learning

To learn an effective representation for each post, we deploy a feature learning layer using the cutting-edge pre-trained language model BERT, which achieves state-of-the-art performance in multiple NLP tasks (Devlin et al., 2019). Since the domain-specific Mental BERT only support English text (Ji et al., 2022) while our annotated testbed is Chinese text, we select the general pre-trained model BERT-base-Chinese to build the feature learning layer. The representation of the [CLS] token in the last layer of BERT can be used as an embedding vector for each post. Specifically, for a post $X_i$, the feature learning layer maps it into an embedding vector:

$$H_i = BERT(X_i) \tag{1}$$

where $H_i$ represents the semantic meaning of a user's $i$-th post $X_i$, and the similarity between $H_i$ and $H_j$ can represent the similarity between post $X_i$ and post $X_j$.

## Temporal Segmentation

Social media data have different levels of granularity. In Figure 5, many studies explore post-level and user-level analysis but largely neglect the potential of period-level analysis.

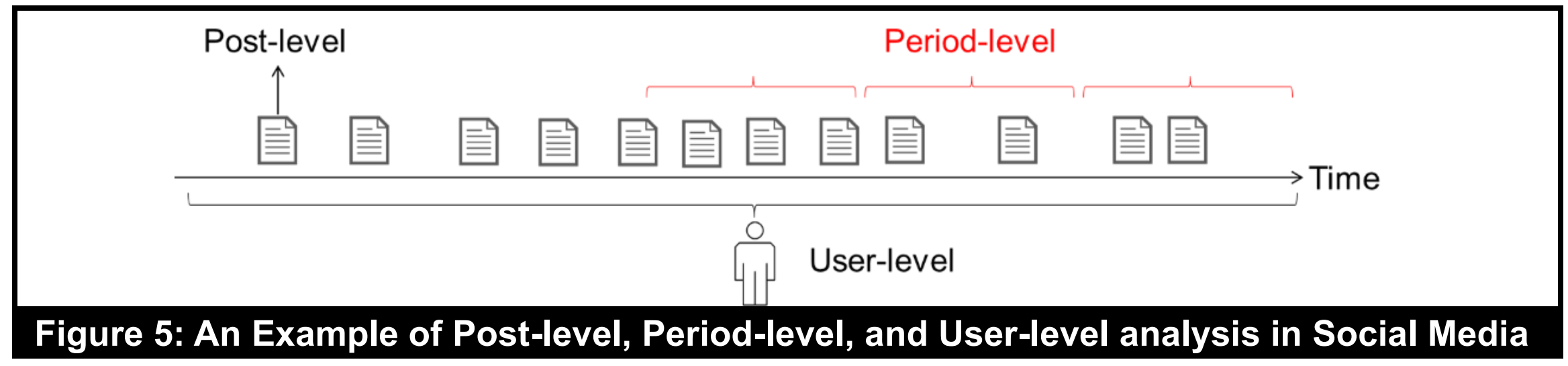

Figure 5: An Example of Post-level, Period-level, and User-level analysis in Social Media

A period refers to a temporal sequence of posts within a pre-defined time window. The period-level analysis aims to infer whether users suffer from depression in a period. Compared to post-level analysis, period-level analysis has two benefits for user-level depression detection in social media. First, given that users may generate repetitive posts in a single depressive episode, the post-level analysis that independently processes each post may overestimate the frequency of depressive episodes, which impedes the predictive power of user-level depression detection. To solve this limitation, the period-level analysis concurrently deals with multiple posts in the same period and uncovers the depression status in a period rather than in a post. Second, different from the post-level static analysis, the period-level analysis has a temporal attribute, which facilitates the measurement of how long depressive symptoms last. Moreover, the period-level analysis aligns with the clinical depression diagnosis criterion that examines patients' depressive symptoms over periods and improves the interpretability of depression detection in social media. Therefore, we use period-level analysis as the intermediate for subsequent user-level detection.

The key to the period-level analysis is the length and position of the period. Given the persistence of depressive symptoms, it is reasonable to assume that the symptoms disclosed by a single post persist for some time, which is determined by the pre-defined length of periods. The length is fixed because varied length falls short of subsequent measurements for the "how long" aspect of depressive symptoms. When this length is too large or too small, it can overestimate or underestimate the duration of depressive symptoms, resulting in low accuracy. The optimal parameter can be determined by empirical analysis. Given the length of a period, it is equally critical to determine the position of the period. Ideally, the start and end of a period should match the patient's actual depressive episode, or non-depression, which is challenging because the patient's depressive episode cannot be directly observed. As shown in Figure 6, a person may be in a depressive episode or non-depression period, and he or she is more likely to generate depression-related posts in a depressive episode.

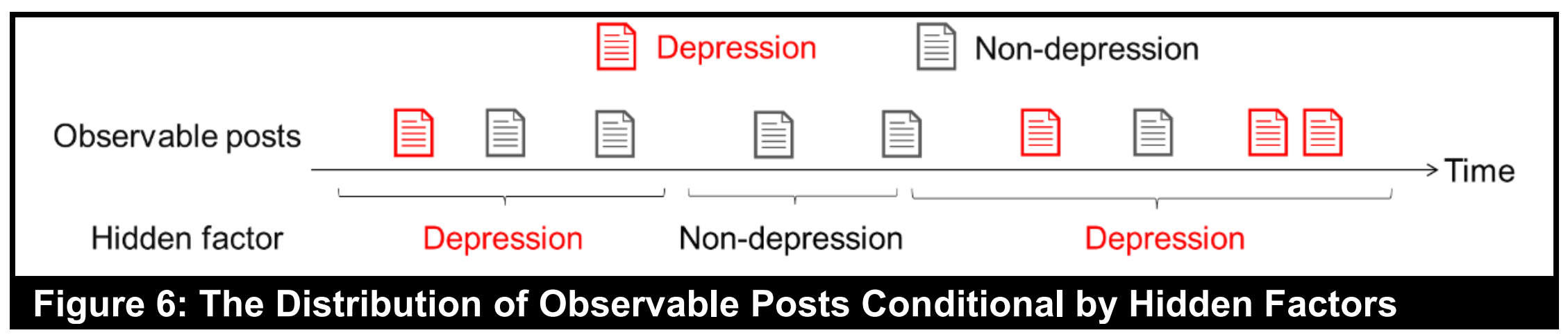

**Figure 6: The Distribution of Observable Posts Conditional by Hidden Factors**

It is non-trivial to segment these observable posts into different periods that respond to the same hidden factor. On one hand, supervised segmentation methods are infeasible for segmenting social media posts due to the difficulty of obtaining sufficient labeled data (Glavaš et al., 2021). On the other hand, there are few unsupervised segmentation methods applied to social media posts. Prior literature on sequential data segmentation focuses on two approaches. One is to segment a sequential text document into different paragraphs based on topic similarity (Riedl & Biemann, 2012). Another is segmenting a sequence of sensor signals into different snippets with a fixed length in order (Yu et al., 2023). Leveraging the advantage of each approach, we propose a temporal segmentation method, which considers both semantic similarity and the time interval between posts. The intuition of our segmentation approach is that semantically similar posts that are generated at adjacent times originate from the same hidden factor.

Our temporal segmentation layer builds upon a bottom-up hierarchical clustering algorithm (Shetty & Singh, 2021) to segment the social media posts $u = (H_1, t_1; H_2, t_2; \dots; H_n, t_n)$ into $m$ periods $u = (C_1, C_2, \dots, C_m)$, where $C_i = (H_{i,1}, t_{i,1}; H_{i,2}, t_{i,2}; \dots; H_{i,l}, t_{i,l})$. Compared to k-means clustering methods with a fixed number of clusters, this algorithm can segment a varied number of clusters (i.e., periods) for each user, which is suitable for our study. This algorithm starts with all posts where each post is considered as one segment and iteratively merges the closest segments until the stop criterion are met, e.g., the maximum length of a period. Let $h$ denote the fixed length of the period. If we set $h$ to 7 days, posts in each period have to be within a week.

The key to segmentation methods is the distance measurement between different posts. Traditional measurements largely rely on the semantic meaning of posts (Huang et al., 2022), neglecting the posting time. In response to this, we propose a new measurement that combines both semantic similarity and the time interval between posts. Specifically, the semantic similarity $sim_{sem}^{i,j}$ of post $i$ and post $j$ is measured by the cosine similarity of their embedding vectors $H_i$ and $H_j$ as shown in Formula (2). The temporal distance $dis_{time}^{i,j}$ of post $i$ and post $j$ is measured by the absolute difference of their posting time $t_i$ and $t_j$ as shown in Formula (3). To fuse the above measurements, we need to normalize the temporal distance between posts using a monotonic decreasing function. The function should be able to adjust the speed of decreasing because the duration of episodes of mental illness such as depression varies. Therefore, we use the Formula (4) to normalize the temporal distance (Kundu et al., 2021), where $w_d$ is a hyperparameter that affects the speed of decreasing. The greater the $w_d$, the smaller the penalty for distant posts. For example, if $w_d$ is 7, then the $sim_{time}^{i,j}$ is 1 when $dis_{time}^{i,j}$ is 0 and is 0.37 when $dis_{time}^{i,j}$ is 7. Given that value ranges of $sim_{time}^{i,j}$ and $sim_{sem}^{i,j}$ are both between 0 and 1, we employ a linear interpolation function to combine them, as shown in Formula (5), where $w_a$ is a weight hyperparameter between 0 and 1. The greater $w_a$, the more important the $sim_{time}^{i,j}$.

$$sim_{sem}^{i,j} = \frac{H_i * H_j}{\|H_i\| * \|H_j\|} \tag{2}$$

$$dis_{time}^{i,j} = |t_i - t_j| \tag{3}$$

$$sim_{time}^{i,j} = \exp\left(-\frac{dis_{time}^{i,j}}{w_d}\right) \tag{4}$$

$$sim^{i,j} = w_a * sim_{time}^{i,j} + (1 - w_a) * sim_{sem}^{i,j} \quad (5)$$

Different from the measurement of similarity between posts in Formula (5), the similarity between clusters (e.g., $C_a$ and $C_b$) needs to consider all posts contained in each cluster. The popular measurements include single linkage, complete linkage, and average linkage. To group as many depression-related posts as possible, we select the single linkage measure, which takes the distance between the two closest posts from two clusters as the distance between the two clusters, as shown in Formula (6). While this measurement may introduce more depression-unrelated posts into each cluster, this noise can be reduced through the next layer of our methods. In each iteration, the method calculates the similarity between each pair of segments (or clusters), and then merges the most similar pair into a new segment, until the time distance between the two segments exceeds the pre-defined length $h$ of periods. The remaining clusters $(C_1, C_2, \ldots, C_m)$ are the segmentation results, where $C_i$ is the $i$-th segment of the focal user, and $X_{i,j}$ is the $j$-th post in the segment $C_i$.

$$Sim^{C_a,C_b} = \min_{i \in C_a, j \in C_b} sim^{i,j} \quad (6)$$

After segmenting, each period contains one or more posts, which jointly present the depressive symptoms of the user in the period. For example, $C_i$ contains two posts disclosing disturbed sleep and one post disclosing suicidal thoughts. This suggests that the user suffered from disturbed sleep and suicidal thoughts in the $i$-th period. Such period-level analysis facilitates the subsequent measurement of how long depressive symptoms last articulated in the next subsection.

**Multi-Scale Temporal Prototype**

The design of this layer is inspired by prototype learning, which draws conclusions for new inputs by comparing them with a few exemplar cases in the problem domain. The proper definition of the prototype is crucial for the final prediction and interpretation. In general, the prototype should be easy for humans to understand and has sufficient corresponding raw data in the training data set to facilitate learning. For example, to classify the sentiment of Yelp reviews, Ming et al. (2019) propose the ProSeNet model that defines prototypes as typical reviews expressing positive or negative sentiment. Different from the simple text classification task, Chen et al. (2019) classify images using the ProtoPNet model, which defines the prototype as part of the image (e.g., the head of a clay-colored sparrow) rather than the whole image (e.g., a clay-colored sparrow) because image data is complex and lacks sufficient representative images in training data. Similarly, typical video clips that look fake are recognized as prototypes to detect fake videos using DPNet proposed by Trinh et al. (2021). When applying these methods in user-level depression detection, the definitions of prototypes would be a typical depressed user, a typical post that discloses depressive symptoms, or typical posts within a period, which uncover typical depressive symptoms. Following the clinical depression diagnosis criterion, our model aims to capture what symptoms users suffer and how long these symptoms last, which necessitates the accurate definition of the prototype. However, the above definitions cannot suffice our research for the following reasons:

(1) It is infeasible to define the prototype as a typical depressed user like ProSeNet because social media users contain many posts, which are difficult for humans to understand at a glance.

(2) While ProtoPNet can infer depressive symptoms from a depression-related post by comparing learned symptom prototypes in our study, these prototypes are static and thus fall short of capturing the “how long” aspect of depressive symptoms.

(3) While DPNet (Trinh et al., 2021) has the potential to capture typical dynamic depressive symptoms from user-generated posts in a period, it still faces two limitations. First, similar to a clip of a video, this model takes all posts in a period as a whole to learn dynamic prototypes but fails to reduce noise from depression-unrelated posts. The problem is similar to randomly inserting unrelated videos into a clip of a video. Second, this model only captures dynamic depressive symptoms that last for one period and fails to detect for longer periods. Although it is a plausible solution to increase the pre-defined length of periods, this also increases the learning complexity, and the learned prototype is difficult for humans to understand.

We propose the novel multi-scale temporal prototype layer to solve the above limitations. This layer first infers what symptoms the user suffers from in each period, then explicitly measure the temporal distribution (e.g., frequency and persistence) of depressive symptoms over periods. Different from taking all posts in a period as a whole to capture depressive symptoms in the period, our method represents depressive symptoms in each period by using a post that is most relevant to a specific depressive symptom. The method that independently analyzing each post in a period and then determining the most representative post can avoid interference from depression-unrelated posts, which contributes to capturing depressive symptoms in each period.

To capture the dynamic depressive symptoms over a longer period, we explicitly measure the temporal distribution of depressive symptoms over continuous periods, rather than increasing the length of the period. The frequency (e.g., I suffer insomnia three times this month) and persistence (e.g., I have lasting insomnia for 7 days) of depressive symptoms are important measurements for depression detection and provide clinically relevant explanations. The frequency tends to measure long-term depressive symptoms and the persistence focuses on the short term. In essence, these measures are differences in time scales, which refer to the number of analyzed continuous periods. To capture a comprehensive temporal distribution of depressive symptoms over periods, inspired by CNN's use of filters with different sizes to extract local features of images, we measure the global and local distribution of depressive symptoms over periods by using different time scales. For example, when the scale is 3 and the target symptom is disturbed sleep, we traverse the pairs of all three consecutive periods and calculate the existence of the symptoms in each pair. Finally, we take the highest value to measure the presence of disturbed sleep at the scale, which measures persistence. The selection of the highest value is because such extreme value can well distinguish depressed users from non-depressed.

Specifically, we define $k$ prototypes $P = (p_1, p_2, \ldots, p_k)$ to be leaned, where each prototype is learnable parameters with the same length as the latent representation of each post. After adequate training, these parameters converge to a latent representation of a typical depressive symptom, such as a loss of interest (IAG, 2011). We can assign $p_i$ with the closest post in the training data to translate prototypes and make them interpretable (Trinh et al., 2021). We will articulate the training and prototype projection later. Next, based on the learned symptom prototypes, we infer the existence $s_{m,k}$ of the symptom $k$ in period $m$ that contains $l$ posts, as shown in Formulas (7), (8) and (9). The $s_{m,j,k}$ denotes the similarity between depressive symptom prototype $p_k$ and latent representation $H_{m,j}$ of the $j$-th post in $m$-th period $C_m$ by using L2 distance (Ming et al., 2019).

$$s_{m,k} = \max_{j=1,2,\ldots,l} s_{m,j,k} \tag{7}$$

$$s_{m,j,k} = exp\left(-d_{m,j,k}\right) \tag{8}$$

$$d_{m,j,k} = \left\|H_{m,j} - p_k\right\|^2 \tag{9}$$

where zero values of $s_{m,l,k}$ can be interpreted as $H_{m,j}$ being completely different from the prototype vector $p_k$, and one means they are identical. The use of the maximum of similarity $s_{m,l,k}$ has two benefits. First, when a period contains multiple posts disclosing the same symptom, only one is retained to avoid the influence of repetitive information. Second, using the maximum value rather than the mean value avoids the non-depression posts diluting useful information. The $s_{m,k}$ provide period-level clues on symptoms and is the basis for subsequent temporal analysis in longer periods. To clearly illustrate our innovation in prototype design, we contrast three designs of prototype learning methods mentioned above based on the similarity calculation between prototypes and corresponding raw data, as shown in the following:

$$\text{MSTPNet (Ours): } s_{m,k} = \max_{j=1,2,\ldots,l} exp\left(-\left\|H_{m,j} - p_k\right\|^2\right) \quad (10)$$

$$\text{ProSeNet: } s'_{u,k} = exp\left(-\|encoder(u) - p'_k\|^2\right) \quad (11)$$

$$\text{ProtoPNet: } s''_{i,k} = exp(-\|H_i - p''_k\|^2) \quad (12)$$

$$\text{DPNet: } s'''_{m,k} = exp\left(-\left\|encoder\left([H_{m,1}, H_{m,2}, \ldots, H_{m,l}]\right) - p'''_k\right\|^2\right) \quad (13)$$

where Formula (10) is from our method, which merges Formulas (7), (8), and (9). The $s'_{u,k}$ aims to capture the similarity between the user $u$ and the $k$-th user-level typical patient prototype $p'_k$ (Ming et al., 2019), as shown in Formula (11). The $s''_{i,k}$ infers the existence of the $k$-th post-level typical symptom prototype $p''_k$ in the $i$-th post of a user (Chen et al., 2019), as shown in Formula (12). The above two methods are not period-level, thus fall short of measuring how long these symptoms last, which is critical for depression detection. Although the $s'''_{m,k}$ shows the presence of the $k$-th typical dynamic depressive symptoms prototype $p'''_k$ in the $m$-th period (Trinh et al., 2021), the measurement of $s'''_{m,k}$ may be inaccurate because it may be inferred by depression-unrelated posts in $[H_{m,1}, H_{m,2}, \ldots, H_{m,l}]$, as shown in Formula (13). Compared to Formula (13), our innovation is to extract the closest post with symptom prototypes to represent a period.

As shown in Figure 7, the $(s_{1,k}, s_{2,k}, \ldots s_{m,k})$ denotes the distribution of symptom $k$ over periods, and $(s_{m,1}, s_{m,2}, \ldots s_{m,k})$ informs what symptoms the user suffers from in the period $k$. Based on the temporal distribution, previous prototype learning methods generally only focus on whether a prototype appears and the frequency of the prototype, i.e., $\max_{i=1,2,\ldots,m} s_{i,k}$ and $\frac{1}{m}\sum_{i=1,2,\ldots,m} s_{i,k}$. However, the above measurements face two limitations. First, these measurements neglect the persistence of depressive symptoms such as persistent disturbed sleep in Figure 7. Second, the measurement of frequency using the proportion of periods where the symptom appears may not be robust. For example, a user has only recently suffered from depression, so only a small percentage of social media posts that disclosing depression.

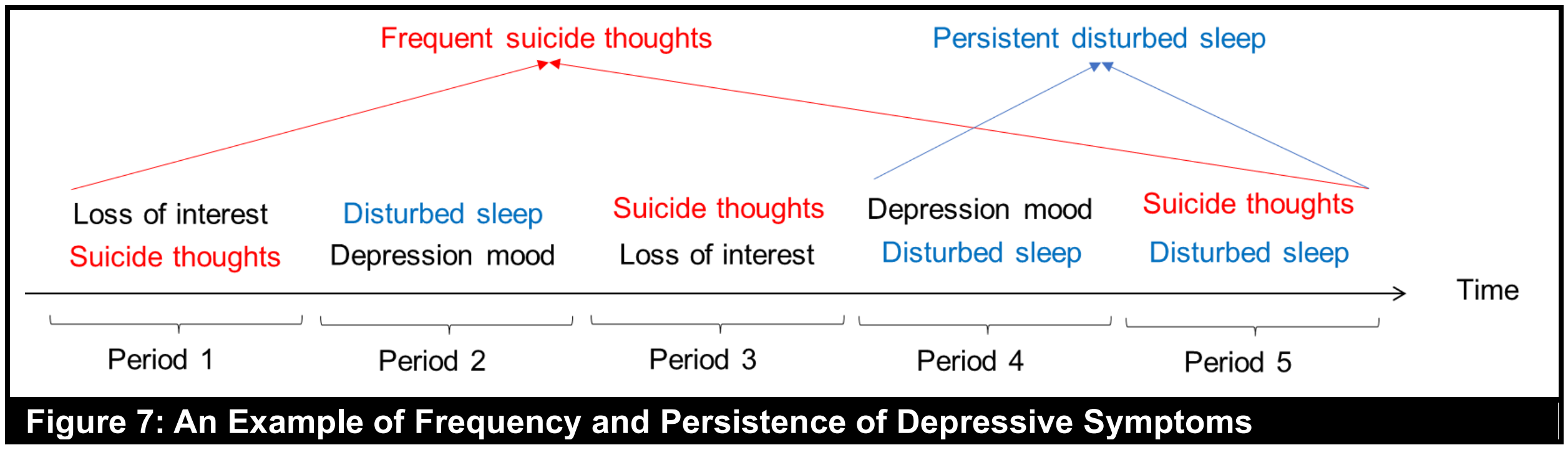

**Figure 7: An Example of Frequency and Persistence of Depressive Symptoms**

Beyond one period or all periods, we focus on a specified number of consecutive periods to measure the frequency and persistence of depressive symptoms. The number of consecutive periods is called “scale” in this study. Different scales enable analysis at different granularity and can capture comprehensive clues to detect depression. When the scale is small, the analysis mainly measures the persistence of depressive symptoms, such as persistent disturbed sleep for two periods in Figure 7. When the scale is large, the analysis focuses on measuring the frequency of depressive symptoms and is insensitive to the duration of a single episode, such as frequent suicide thoughts in Figure 7. Moreover, a scale that is smaller than the total number of periods can alleviate the cold start problem of the newly depressed user because the relatively smaller scale can focus on the recent periods and reduce the negative effect of early periods. Therefore, we employ multiple scales (i.e., multi-scale) with different sizes for period-level analysis, which is conceptually similar to the filters with different sizes in CNN to analyze image data. The difference is that our “filters” are not learnable parameters but are explicitly set to get the average value over continuous periods (e.g., [0.5, 0.5] for scale size = 2), which is easy for humans to understand. Specifically, let $W$ be the set of scales used in our model, and $w_j$ denotes the size of the $j$-th scale. We calculate the existence (i.e., average similarity) of depressive symptoms in each pair with a window length of the scale, and then take the highest value as the existence (i.e., $g_{j,k}$) of the depressive symptom $k$ on the scale $w_j$, as shown in Formula (14).

$$g_{j,k} = \max_{m=1,2,\ldots,M-w_j+1} \frac{1}{w_j} \sum_{m}^{m+w_j-1} s_{m,k} \tag{14}$$

The multi-scale temporal prototype layer is our main methodological contribution. This layer can effectively capture what symptoms users suffer and how long these symptoms last by carefully setting different scales. Although setting too many scales seem to output too many $g_{j,k}$ to increase the burden of end users, the problem can be addressed by showing a few important $g_{j,k}$ (Ribeiro et al., 2016). This layer explicitly separates the task of identifying complex prototypes into two relatively simple tasks, which enhances cost-effectiveness and flexibility.

## Classification

Considering $g_{j,k}$ at multiple scales, the temporal distribution of symptom $k$ can be well captured, including long-term frequency and short-term persistence, which can be used to detect depression. Let $G = (g_{1,1}, g_{1,2}, \ldots g_{1,k}; g_{2,1}, \ldots g_{v,k})$, where $v$ is the number of scales. The classification layer computes the probability of depression given $G$ of a focal user, as shown in Formulas (15) and (16):

$$\hat{y}_i = \frac{\exp(z_i)}{\sum_{s=0}^{1} \exp(z_s)} \tag{15}$$

$$Z = QG \tag{16}$$

where $Q$ is a $2 \times (v \times k)$ weight matrix, $G$ is a $(v \times k) \times 1$ weight matrix. The classification weights $Q$ indicate which symptom prototype at which scale is more important for classification. To enhance interpretability, we constrain $Q$ to be nonnegative (Ming et al., 2019). As shown in Formula (16), $Z$ is a vector with length 2, which represents the total score of each class ($i = 1$ refers to the depression class). SoftMax is used to compute the output probability in Formulas (15), where $z_i$ and $\hat{y}_i$ is the score and probability that the user is classified into the $i$-th class.

Following Ming et al. (2019), the loss function of MSTPNet to be minimized is defined based on the binary cross-entropy (CE) loss with four additional regularization terms.

$$Loss = CE + \lambda_c R_c + \lambda_e R_e + \lambda_d R_d + \lambda_{l_1} \|Q\|_1 \quad (17)$$

$$CE = \sum_{(X,y)\in D} y \log \hat{y} + (1-y)\log(1-\hat{y}) \quad (18)$$

where $\lambda_c$, $\lambda_e$, $\lambda_d$ and $\lambda_{l_1}$ are hyperparameters that determine the weight of the regularizations. The clustering regularization $R_c$ encourages a clustering structure by minimizing the distance between an input and its closest prototype. $R_e$ is an evidence regularization, which encourages each prototype to be as similar as possible to real input. The diversity regularization $R_d$ penalizes prototypes that are close to each other, where $d_{min}$ is a threshold that determines the minimum distance to be included in the calculation. The $\|Q\|_1$ is $L_1$ sparsity penalty. The configuration of these hyperparameters largely depends on the nature of the data and can be selected through cross-validation. We will finetune the hyperparameter settings in empirical analysis.

## Reasoning Process of MSTPNet's Interpretation

To translate the prototypes $P = (p_1, p_2, \dots, p_k)$ as interpretable information, we project each prototype $p_j$ onto a nearest user-generated post based on the distance of $p_j$ and latent representation of posts. In this way, we can conceptually visualize each prototype with a typical post that discloses depressive symptoms. For $p_j$, we can find its corresponding post $seq_j$:

$$seq_j = \underset{x\in TD}{\arg\min} \left\| BERT(x) - p_j \right\|_2 \quad (19)$$

where $TD$ refers to the training dataset, and $x$ is a post from $TD$. The $BERT(x)$ is an embedding vector with equal length as $p_j$. The $x$ with the minimum distance from $p_j$ is the prototype's interpretation, which guides us to associate each prototype with a depressive symptom.

Figure 8 shows the reasoning process of our MSTPNet's interpretation of a test user. The user posted on multiple topics, including depression and non-depression content. Our method firstly considers the time and semantic meaning of posts to segment similar posts into a period. In this example, we set the length of the period to 7 days. That is, the maximum time interval between posts in each period is no more than one week. In Figure 8, we show the learned prototypes and their closest original post, where the corresponding symptoms are shown in parentheses. Then, we infer the presence of depressive symptoms in each period by comparing these learned symptom prototypes with posts in each period. For example, the first learned symptom prototype of the red box "I took sleeping pills, what the hell, why did I wake up?" belongs to the disturbed sleep. A representative post "every morning, I wake up with pain and discomfort" in week 1, has a similarity with 0.45 to the symptom prototype of disturbed sleep, which suggests that the user may suffer from disturbed sleep symptoms in week 1. Similarly, we also can infer the temporal distribution of other depressive symptoms over periods. For example, the existence strengths of loss of interest or pleasure symptoms over weeks 2, 3, and 4 are 0.65, 0.55, and 0.45.

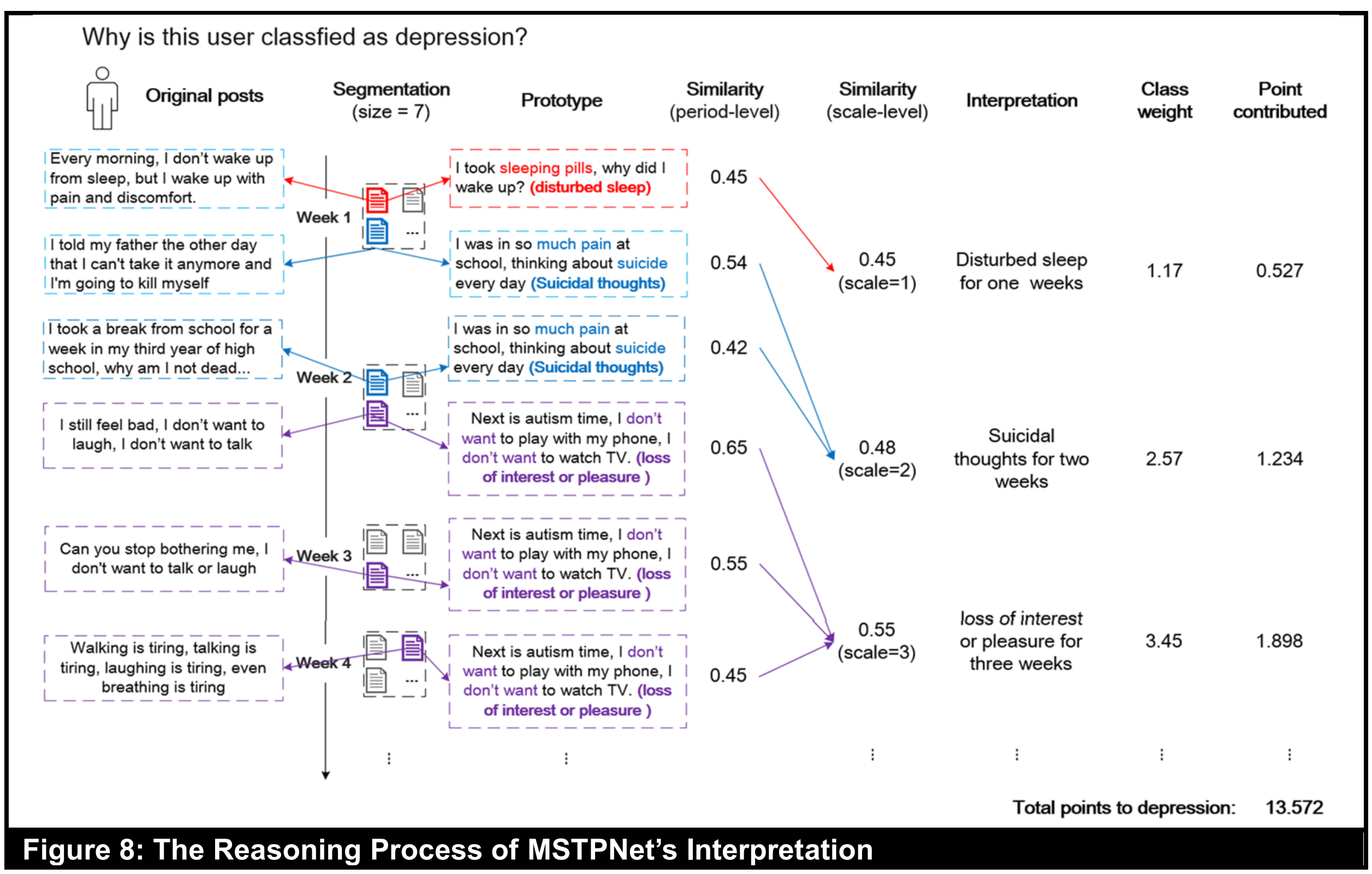

**Figure 8: The Reasoning Process of MSTPNet's Interpretation**

Next, the multi-scale temporal prototype captures the frequency and persistence of depressive symptoms at different time scales. The scale size determines the number of analyzed consecutive periods and affects the level of granularity of the information obtained. For example, when the time scale is 2, the depressive symptoms in any two consecutive weeks are considered, and such short periods focus on the persistence of depressive symptoms. In Figure 8, week 1 and week 2, both contain a post that has a high similarity (i.e., 0.54 and 0.42) with the prototype of suicidal thoughts “I am in such pain at school and think about suicide every day.” The mean value is 0.48, which is also relatively high, indicating that the user experienced suicidal thoughts for two consecutive weeks. The linear weights of the classification layer indicate the importance of the existence strength of depressive symptoms at different scales. Combining the computed similarity and weights yields output scores for the depression and non-depression categories. The category with the highest score is the output.

## EMPIRICAL ANALYSIS

### Data Collection and Preprocess

A large, publicly available dataset, the WU3D dataset (Wang et al., 2022), is used as the research test bed. The WU3D is an annotated dataset regarding depression detection collected from Sina Weibo, which is the most popular social media platform in China. The dataset contains chronological sequences of user-generated posts from 10,325 depressed users and a random control group of 22,245 users. Depressed users are identified by inviting professional data labelers to complete the data labeling process. Moreover, the labeled data has been reviewed twice by psychologists and psychiatrists, using the DSM-5 as the reference. Any occurrence of usernames has been replaced to anonymize users. Table 6 summarizes the main statistics.

**Table 6: Main Statistics of WU3D Dataset**

| Category | User | Post | Posts per user | Words per post |
|---|---|---|---|---|
| Depression | 10,325 | 408,797 | 39.6 | 91.3 |
| Non-depression | 22,245 | 1,783,113 | 80.2 | 47.4 |
| Total | 32,570 | 2,191,910 | 67.3 | 55.6 |

To prepare our experimental data, we first remove posts that are not original or with limited words because they are usually just emojis and stop-words. We also remove inactive users whose historical posts are fewer than ten. In the previous literature, datasets are frequently constructed as balanced (Chau et al., 2020; Zogan et al., 2022). However, there are more healthy people than depressed patients in the real world. To rigorously evaluate our proposed method, we set the ratio between depression and non-depression as 1:8, which approximates the ratio of adults with depression risk according to the report on National Mental Health Development in China (2021-2022) (Fu et al., 2023). We set the sample size as 15,000. Our data size is in line with or larger than most social media-based depression detection studies (Chau et al., 2020; Trotzek et al., 2020). We split this dataset into 60% for training, 20% for validation, and 20% for testing.

## Experiment Design

Following the computational design science paradigm, we extensively evaluate the predictive power and interpretability of our proposed MSTPNet for user-level depression detection in social media. To rigorously validate MSTPNet's predictive power, we compare MSTPNet against state-of-the-art benchmark models, analyze the effect of various hyperparameters on MSTPNet's performance, and perform ablation studies to show the effectiveness of multiple novel design components. These benchmark models include traditional machine learning models, black-box deep learning models, and interpretable deep learning models. For traditional machine learning models, we compile the most common features such as LIWC and N-grams from the social media-based depression detection literature (Liu et al., 2022). The recent IS study (Chau et al., 2020) on depression detection, which combines feature-based and rule-based, is also included. For black-box deep learning models, we compare CNNs, RNNs, attention-based networks, and their combinations and variants, which have state-of-the-art predictive power but lack interpretability (Malhotra & Jindal, 2022). The closest to our study is interpretable deep learning methods, which simultaneously preserve or even improve the predictive power of depression detection and provide interpretation. We compare the predictive power of representative models, including attention-based and symptom-based (Adarsh et al., 2023; Zhang et al., 2022b). Since the approximation-based method is model-agnostic, we only compare it in terms of interpretability. In addition, we contrast three popular prototype learning methods such as ProSeNet, ProtoPNet, and DPNet, which are close to our method and can be adapted to interpretable depression detection in social media (Li et al., 2022). A detailed description of the benchmark models can be found in Table 7.

**Table 7: Benchmark Models**

| Category | | Literature | Models | Description |
|---|---|---|---|---|
| Traditional Machine Learning | | Choudhury et al. (2013) | SVM | The inputs are the most common features such as LIWC and N-grams from the literature. |
| | | Yang et al. (2020) | LR | |
| | | Chen et al. (2018) | RF | |
| | | Chiong et al. (2021b) | DT | |
| | | Chau et al. (2020) | SVM + Rule-based | Combines feature-based and rule-based |
| Black-Box Deep Learning | | Orabi et al. (2018) | CNN | Leverage pre-trained models to encode text words or posts. The main input is the content of the posts, and few models consider the posting time. |
| | | Chiu et al. (2021) | LSTM with temporal weighting and day-based aggregations | |
| | | Ghosh and Anwar (2021) | LSTM | |
| | | Naseem et al. (2022) | Bi-LSTM + Attention | |
| | | Kour and Gupta (2022) | CNN + Bi-LSTM | |
| Interpretable Deep Learning | Approximation-based | Adarsh et al. (2023) | Black-box model + LIME | Ensemble SVM and KNN. Post-hoc LIME provides the importance of each feature or input elements |
| | Attention-based | Cheng and Chen (2022) | Time-aware attention network | Inherently interpretable. The attention weight highlights important input elements. |
| | | Zogan et al. (2022) | HAN | |
| | Symptom-based | Zhang et al. (2022a) | Symptom Template + HAN | Screen out important posts based on symptom templates |
| | | Ming et al. (2019) | ProSeNet | For text classification |
| | | Chen et al. (2019) | ProtoPNet | For image classification using the typical part of the image |
| | | Trinh et al. (2021) | DPNet | For time-series classification |

Sensitivity analysis is performed to evaluate how the selection of hyperparameters affects MSTPNet's predictive power. To avoid exponential numbers of hyperparameter combinations, we use our proposed MSTPNet as a baseline and adjust only one type of hyperparameter at a time. A detailed description of the sensitivity analysis can be found in Table 8.

**Table 8: Sensitivity Analysis**

| Hyperparameter | Options | Model name | Hyperparameter | Options | Model name |
|---|---|---|---|---|---|
| Length of Periods | 3 | MSTP-S (3) | Number of Prototypes | 30 | MSTP-N (30) |
| | 5 | MSTP-S (5) | | 40 | MSTP-N (40) |
| | 7 | MSTP-S (7) | | 50 | MSTP-N (50) |
| | 9 | MSTP-S (9) | | 60 | MSTP-N (60) |
| | 11 | MSTP-S (11) | | 70 | MSTP-N (70) |
| | 13 | MSTP-S (13) | | 80 | MSTP-N (80) |
| | 15 | MSTP-S (15) | | 90 | MSTP-N (90) |
| | 17 | MSTP-S (17) | | 100 | MSTP-N (100) |
| Weight of temporal distance | 0 | MSTP-W (0) | Combination of Scale Size | 1 | MSTP-C (1) |
| | 0.1 | MSTP-W (0.1) | | 1, 2 | MSTP-C (2) |
| | 0.2 | MSTP-W (0.2) | | 1, 2, 3 | MSTP-C (3) |
| | 0.3 | MSTP-W (0.3) | | 1, 2, 3, 5 | MSTP-C (5) |
| | 0.4 | MSTP-W (0.4) | | 1, 2, 3, 5, 8 | MSTP-C (8) |
| | 0.5 | MSTP-W (0.5) | | 1, 2, 3, 5, 8, 12 | MSTP-C (12) |
| | 0.6 | MSTP-W (0.6) | | 1, 2, 3, 5, 8, 12, 16 | MSTP-C (16) |
| | 0.7 | MSTP-W (0.7) | | 1, 2, 3, 5, 8, 12, 16, 20 | MSTP-C (20) |
| | 0.8 | MSTP-W (0.8) | | | |
| | 0.9 | MSTP-W (0.9) | | | |
| | 1 | MSTP-W (1) | | | |

To understand the interpretability of our model, we will show the learned symptom prototypes and analyze their importance at different scales. To validate the superiority of MSTPNet in terms of interpretability for social media-based depression detection, we first visualize and contrast different forms of interpretation, including approximation-based, attention-based, symptom-based, and our proposed symptom-based with how long these symptoms. Then we conduct a user study to compare users' trust and perceived helpfulness in our model and alternative models.

## Evaluation of Predictive Power

### Comparison against Benchmark Models

We adopt F1-score, precision, recall, and accuracy as the evaluation metrics. The best model should have the highest F1 score. The evaluation results are reported in Tables 9, 10, and 11.

These metrics are the mean of 10 experimental runs. MSTPNet outperforms benchmark models in F1 score and accuracy and outperforms interpretable deep learning models in all metrics.

| Table 9: Predictive Power of Depression Detection against Traditional Machine Learning Models | | | | |
|---|---|---|---|---|
| Models | F1 | Precision | Recall | Accuracy |
| Choudhury et al. (2013) | 0.665*** | 0.861*** | 0.542*** | 0.942*** |
| Yang et al. (2020) | 0.412*** | 0.887*** | 0.268*** | 0.918*** |
| Chen et al. (2018) | 0.508*** | 0.891*** | 0.355*** | 0.926*** |
| Chiong et al. (2021b) | 0.380*** | 0.363*** | 0.399*** | 0.861*** |
| Chau et al. (2020) | 0.706*** | 0.604*** | 0.850 | 0.924*** |
| **MSTPNet** | **0.851** | **0.957** | **0.766** | **0.971** |

Note: *p < 0.05; **p < 0.01; ***p < 0.001

As shown in Table 9, Compared to the best-performing traditional machine learning model (Chau et al., 2020), MSTPNet improves F1-score by 0.145. The improved performance shows the advantage of deep learning's automatic feature extraction without manual design.

| Table 10: Predictive Power of Depression Detection against Black-box Deep Learning Models | | | | |
|---|---|---|---|---|
| Models | F1 | Precision | Recall | Accuracy |
| Orabi et al. (2018) | 0.828*** | 0.878*** | 0.785 | 0.965** |
| Chiu et al. (2021) | 0.822** | 0.936* | 0.732** | 0.966** |
| Ghosh and Anwar (2021) | 0.820** | 0.921** | 0.741*** | 0.965*** |
| Naseem et al. (2022) | 0.826** | 0.963 | 0.723** | 0.967* |
| Kour and Gupta (2022) | 0.781*** | 0.836*** | 0.732** | 0.956** |
| **MSTPNet** | **0.851** | **0.957** | **0.766** | **0.971** |

Note: *p < 0.05; **p < 0.01; ***p < 0.001

Compared to the best-performing black-box deep learning model (Orabi et al., 2018), MSTPNet improves F1-score by 0.023 as shown in Table 10. Such a significant performance gain indicates that the pursuit of interpretability can assist the learning process to pick up useful information and avoid learning noise, thus enhancing the performance of the model.

| Table 11: Predictive Power of Depression Detection against Interpretable Deep Learning Models | | | | |
|---|---|---|---|---|
| **Models** | **F1** | **Precision** | **Recall** | **Accuracy** |
| Cheng and Chen (2022) | 0.806*** | 0.910** | 0.723** | 0.963** |
| Zogan et al. (2022) | 0.795*** | 0.840*** | 0.754* | 0.958** |
| Zhang et al. (2022a) | 0.726*** | 0.863*** | 0.626*** | 0.949*** |
| Ming et al. (2019) | 0.816** | 0.929* | 0.729*** | 0.965** |
| Chen et al. (2019) | 0.735*** | 0.876*** | 0.633*** | 0.951** |
| Trinh et al. (2021) | 0.675*** | 0.774*** | 0.598*** | 0.938*** |
| **MSTPNet** | **0.851** | **0.957** | **0.766** | **0.971** |

Note: *p < 0.05; **p < 0.01; ***p < 0.001

Compared to the best-performing interpretable deep learning model, MSTPNet improves F1-score by 0.035 (Table 11). Such a significant performance gain indicates that considering the temporal distribution of prototypes greatly contributes to depression detection.

**Sensitivity Analysis**

**Table 12: Sensitivity Analysis Results**

| Model | F1 | Precision | Recall | Accuracy |
|---|---|---|---|---|
| MSTP-S (3) | 0.833 | 0.921 | 0.760 | 0.967 |
| MSTP-S (5) | 0.834 | 0.915 | 0.766 | 0.967 |
| MSTP-S (7) | 0.834 | 0.924 | 0.760 | 0.968 |
| MSTP-S (9) | 0.840 | 0.928 | 0.766 | 0.969 |
| MSTP-S (11) | 0.846 | 0.950 | 0.763 | 0.970 |
| MSTP-S (13) | 0.857 | 0.947 | 0.782 | 0.972 |
| **MSTP-S (15)** | **0.851** | **0.957** | **0.766** | **0.971** |
| MSTP-S (17) | 0.834 | 0.952 | 0.741 | 0.968 |
| MSTP-W (0) | 0.771 | 0.909 | 0.671 | 0.957 |
| MSTP-W (0.1) | 0.769 | 0.977 | 0.636 | 0.961 |
| MSTP-W (0.2) | 0.786 | 0.897 | 0.703 | 0.958 |
| MSTP-W (0.3) | 0.834 | 0.944 | 0.749 | 0.968 |
| **MSTP-W (0.4)** | **0.851** | **0.957** | **0.766** | **0.971** |
| MSTP-W (0.5) | 0.811 | 0.949 | 0.710 | 0.965 |
| MSTP-W (0.6) | 0.814 | 0.949 | 0.714 | 0.966 |
| MSTP-W (0.7) | 0.832 | 0.980 | 0.723 | 0.972 |
| MSTP-W (0.8) | 0.794 | 0.972 | 0.673 | 0.964 |
| MSTP-W (0.9) | 0.844 | 0.948 | 0.760 | 0.970 |
| MSTP-W (1) | 0.810 | 0.933 | 0.718 | 0.964 |
| MSTP-N (30) | 0.828 | 0.928 | 0.748 | 0.967 |
| MSTP-N (40) | 0.827 | 0.927 | 0.746 | 0.967 |
| MSTP-N (50) | 0.835 | 0.923 | 0.762 | 0.968 |
| MSTP-N (60) | 0.836 | 0.922 | 0.765 | 0.968 |
| **MSTP-N (70)** | **0.851** | **0.957** | **0.766** | **0.971** |
| MSTP-N (80) | 0.838 | 0.930 | 0.760 | 0.968 |
| MSTP-N (90) | 0.843 | 0.924 | 0.775 | 0.969 |
| MSTP-N (100) | 0.842 | 0.937 | 0.764 | 0.969 |
| MSTP-C (1) | 0.747 | 0.856 | 0.665 | 0.952 |
| MSTP-C (2) | 0.799 | 0.911 | 0.712 | 0.962 |
| MSTP-C (3) | 0.815 | 0.909 | 0.739 | 0.964 |
| MSTP-C (5) | 0.817 | 0.902 | 0.746 | 0.964 |
| MSTP-C (8) | 0.824 | 0.922 | 0.745 | 0.966 |
| MSTP-C (12) | 0.832 | 0.920 | 0.760 | 0.967 |
| MSTP-C (16) | 0.834 | 0.929 | 0.758 | 0.968 |
| **MSTP-C (20)** | **0.851** | **0.957** | **0.766** | **0.971** |

Table 12 summarizes the effect of important hyperparameters on the performance of our proposed MSTPNet. The model names and results in boldface indicate the hyperparameters we choose among the different options. As shown in Table 9, the 13-day or 15-day length periods reach the best performance, which is also in line with clinical practice. In the PHQ-9 depression questionnaire that is used to clinically diagnose depression, the timeframe for mental health observation is two weeks. The table also shows that the optimal weight of temporal distance is 0.4. The results suggest that combining temporal distance and the similarity of semantic meaning between posts is superior to anyone single distance. As the number of prototypes increases, the performance of the model increases, but there is no significant improvement after 70. This is because the number of valid prototypes is limited, and setting too many prototypes can result in some learned prototypes being too similar. To improve interpretability, we choose 70 as the optimal number of the hyperparameter. We also test the effect of scale sizes on MSTPNet. The results show that the comprehensive use of the information obtained from the multi-scale perspective can further improve the predictive power of the model. However, although adding more scales can further improve performance, the selection of scale size and quantity should be determined according to specific conditions and requirements, such as training time and cost. We also need to consider the effects of too much scale on interpretability, although we can show a few important variables to provide interpretations.

**Ablation Studies**

**Table 13: Ablation Studies**

| Model | F1 | Precision | Recall | Accuracy |
|---|---|---|---|---|
| MSTPNet (Ours) | **0.851** | **0.957** | **0.766** | **0.971** |
| MSTPNet removing temporal segmentation layer | 0.801*** | 0.923*** | 0.702*** | 0.962*** |
| MSTPNet removing MS using Max | 0.760*** | 0.868*** | 0.676*** | 0.954*** |
| MSTPNet removing MS using Mean | 0.690*** | 0.846*** | 0.583*** | 0.944*** |

Since our model consists of multiple critical design components, we further perform ablation studies to show their effectiveness. We remove the temporal segmentation layer to validate the effectiveness of period-level analysis. We also replace the multi-scale temporal prototype (MS) layer with a common prototype learning layer. To validate the effect of multi-scale temporal measurements, we test two options: using the maximum existence strength of prototypes and using the mean existence strength of prototypes over periods to detect depression. Table 13 suggests that removing any design component will significantly hamper detection accuracy.

## Evaluation of Interpretability

### Projection of Symptom Prototypes

**Table 14: Interpretation of Symptom Prototypes**

| Interpretation of Symptom Prototypes | ICD-11/**New** |
|---|---|
| Don't conceive when you are emotionally depressed. Once you are anxious, depressed, or have a heavy mental burden. | Depressed mood |
| I can't seem to find anyone to talk to except on Weibo. It's so sad that I can't keep going. No one cares about whether I'm okay or not. No one cares my condition | fatigue or low energy |
| Unhappy, unhappy, too depressed, too depressed. I said I was depressed, and he said, do you want to commit suicide? I don't know if cutting my hands as suicide | Suicidal thoughts |
| Standing at the window and wanting to jump off, I'm really sad, I can't do anything, I'm so tired, I'm counting the time, I'm leaving tonight, I hope there's no pain | Suicidal thoughts |
| Different from typical depression, patients with smiling depression do not stay at home every day and do not socialize with others, but have good social functions | **Share depression-related content** |
| I tossed and turned and still couldn't fall asleep. I took antidepressants and sleeping pills indiscriminately. I slept soundly. I woke up at 12 o'clock at noon. | Disturbed sleep |
| If a woman doesn't get enough sleep, she suffers more mentally and physically than a man, and may increase her risk of heart disease, depression, etc. | **Share depression-related content** |
| A total of 63,593 severe cases were cured and discharged from hospital 46,441 cumulative deaths in Wuhan 4,512 3,869 cumulative confirmed cases in Wuhan | **Share negative news/events** |
| It is also in line with the principle of the non-duality of form and emptiness in the Heart Sutra. This principle reveals that all dharmas in the ten directions and three times in the world are nothing | **Share admiration for different life** |
| First time seeing a psychiatrist in 2019 Depression I haven't told anyone I don't know what to do I've been thinking a lot about failing to live | Low self-confidence |

MSTPNet is capable of interpreting why a user is classified as depressed by presenting what symptoms the user suffers from and how long these related symptoms last. The interpretability mostly relies on the learned symptom prototypes. Since the prototype vectors are representations in the latent space, they are not readily interpretable. Following the widely adopted prototype projection method from previous prototype learning studies (Ming et al., 2019), we can visualize a prototype by a typical post that discloses depressive symptoms with the most similar latent representation. Moreover, to improve the readability of the responding typical post for each prototype, we prefer a short post with high similarity, instead of the post with the highest similarity. Table 14 shows the ten most salient prototypes and their responding posts as interpretations. For clarity, we highlighted the keywords related to symptoms in each post.

For each prototype, we match it with the symptoms defined by ICD-11, such as depressed mood. Table 14 also shows a few depressive symptoms unnoted in previous medical literature, including sharing depression-related content, and sharing admiration for a different life. We also find two prototypes in the third and fourth row of Table 14 for suicidal thoughts symptom, where the prototype focuses on the experience of suicide and the latter focus on the plan of suicide to be executed. Therefore, our MSTPNet can discover new symptoms as well as analyze existing symptoms at a more granular level, which pre-defined symptoms-based methods fall short of.

**Interpretation of How Long Depressive Symptoms Last**

Different from the previous prototype learning methods, our method not only makes prototypes interpretable but also can interpret how long depressive symptoms last. The latter interpretation relies on the existence strength of depressive symptoms at different scales. For example, let the $j$-th scale be 3, the $g_{j,k}$ refer to the maximum likelihood of the $k$-th depressive symptom occurring in three consecutive periods, which mainly measures the persistence of the depressive symptom. However, our model can select multiple scales (e.g., 1, 2, 3, 5, …, 20), which produce so many results that are difficult for the end user to read and understand. To improve clarity, only a few interpretable elements (e.g., $g_{j,k}$) that are most important to the predicted results should be shown (Ribeiro et al., 2016). The importance of interpretable elements relies on the type of depressive symptoms and their scale. We show the ten most important depressive symptoms in Table 14, and we also need to analyze different scales.

As shown in Table 15, we show an example of the weights of the existence strength of depressive symptoms at different scales (i.e., $g_{j,k}$) for depression class. For example, 4.1 in the first column of the table means the importance of the presence of the depressed mood symptom in a single period, while 2.5 means the importance of the appearance of the symptom in two consecutive periods. Traditional prototype learning methods and deep learning methods generally only capture information on either scale 1 or max scale, corresponding to the maximum pooling and average pooling, which overlooks the persistence of symptoms.

**Table 15: An Example of the Weight of Symptom Prototypes with Different Scales**

| 1 | 2 | 3 | 5 | 8 | 12 | 16 | 20 | Symptom Prototype |
|---|---|---|---|---|---|---|---|---|
| 4.1 | 2.5 | 2.0 | 1.8 | 2.1 | 2.9 | 3.3 | 3.1 | Don't conceive when you are emotionally depressed. Once you are anxious, depressed or have a heavy mental burden. |
| 3.9 | 2.0 | 1.5 | 1.2 | 1.4 | 2.1 | 2.5 | 2.4 | I can't seem to find anyone to talk to except on Weibo. It's so sad that I can't keep going. |
| 3.6 | 1.8 | 1.4 | 1.2 | 1.2 | 1.6 | 1.8 | 1.8 | Unhappy, unhappy, too depressed, too depressed. I said I was depressed, and he said, do you want to commit suicide? |

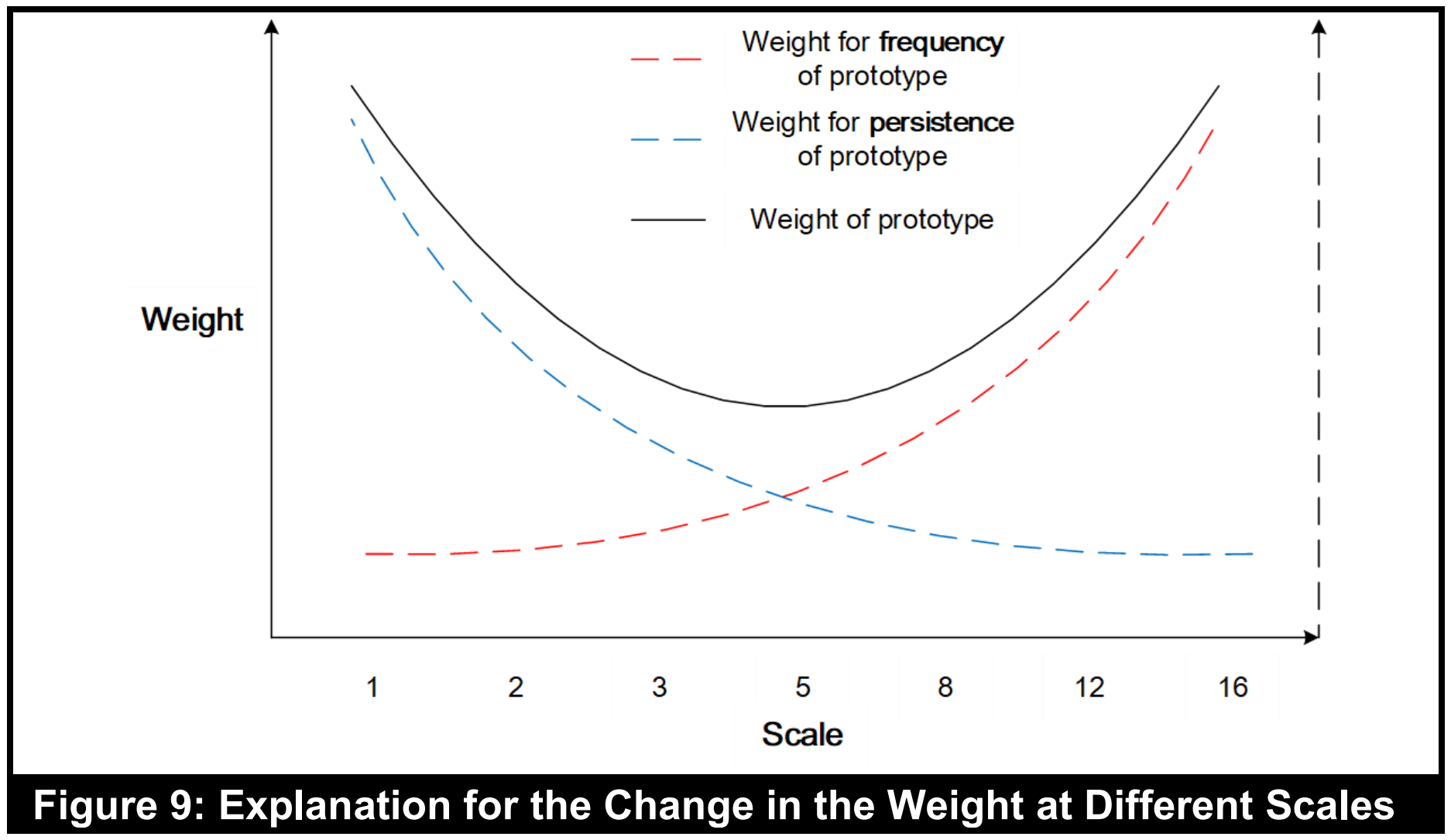

**Figure 9: Explanation for the Change in the Weight at Different Scales**

It is interesting to note that the weight decreases first and then increases when the scale increases in Table 15. We attempt to explain the reason behind the U-shaped weight change of prototypes in Figure 9. When the scale is low, the model mainly considers the persistence of symptoms of single depressive episodes. When the scale is high, the model mainly considers the frequency of symptoms. The above two scales have the highest weight, which is consistent with the fact that we often use the maximum and the average to measure distribution in real life.

When the scale falls in the middle, both long-term and short-term information are considered. Thus, the weight at a medium scale will is a combination of weight for persistence and frequency of the prototype. For example, while the existence strength at a scale = 5 can capture the persistence of depressive symptoms, the importance of information is lower than at a scale < 5. Similarly, while the existence strength at a scale = 5 can capture the frequency of depressive symptoms, the importance of information is lower than at a scale > 5. Since the value of the two types of information captured by scale = 5 is low, the final weight is also the lowest in all scales.

Therefore, when interpreting the temporal distribution of depressive symptoms, we can simply show the existence strength of the depressive symptom on a small scale (not the smallest) and a large scale (not the largest). For example, the reason why a user is classified as depressed is "the user experienced disturbed sleep disorder for three consecutive weeks in June" (persistence), and "the user has a 20% of weeks suffer suicidal thoughts this year" (frequency). The visual interpretation should consider the importance of both symptoms and scales.

**Visual Comparison of Different Types of Model Interpretability**

Our MSTPNet provides a level of interpretability that is absent in other interpretable deep models. In terms of the type of explanations offered, Figure 10 provides a visual comparison of different types of model interpretability. The approximation-based explanation mainly reveals the importance score and direction of interpretable input (e.g., words) linked to the final output, as shown in Figure 10(a). At a finer level, attention-based explanation enables the end user to attend to important words within a post, and important posts for depression detection, as shown in Figure 10(b). The above two explanations are common in many scenarios but fall short of interpreting depression detection because they depart from the clinical depression diagnosis criterion that is based on depressive symptoms. Figure 10(c) shows a symptom-based explanation that contains the strength of each depressive symptom, which is generally measured by the maximum similarities between user-generated posts and each symptom. The posts with the highest similarity are presented as evidence, which provides the end user to judge the credibility of the interpretation.

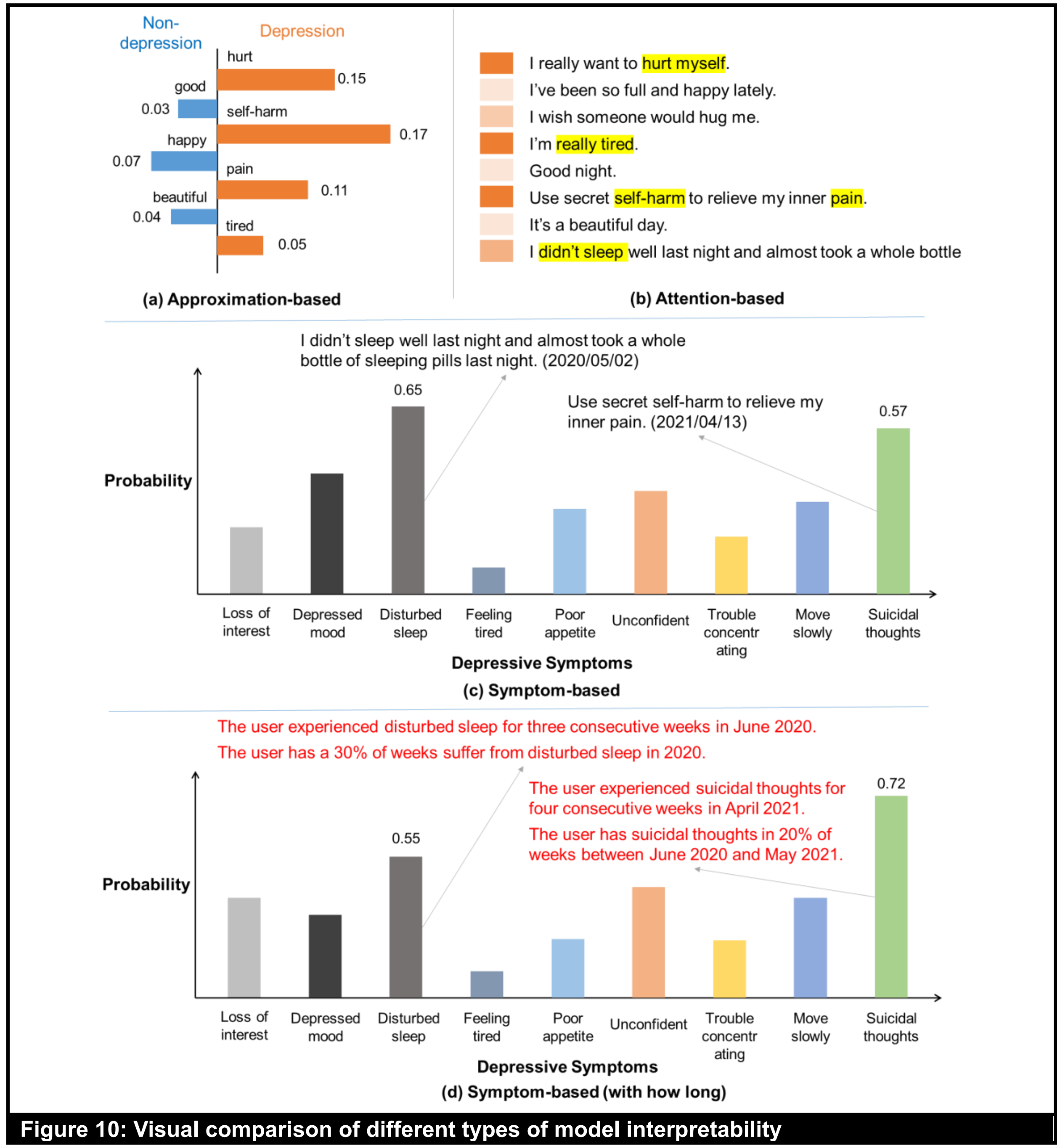

**Figure 10: Visual comparison of different types of model interpretability**

Our interpretation is also symptom-based, but the innovation is that we combine the duration of symptoms to more accurately measure the existence strength of each depressive symptom, as shown in Figure 10(d). The interpretation is capable of capturing what symptoms users suffer and how long these symptoms last, which aligns with the clinical depression diagnosis criterion. Moreover, our MSTPNet is based on prototype learning and can show new depressive symptoms rather than just pre-defined symptoms. In addition to the qualitative comparison, in the next section, we will quantitatively compare our model with the baseline in terms of perceived usefulness and trust, which are major components of interpretability, by way of questionnaires.

**Human Evaluation**

We recruited 92 volunteers and informed the participants that they would be assigned an interpretable ML model to predict depression using social media data. We randomly selected one user sample that the model classified as depressed and show the participants how the model interpreted this classification. Since our main methodological contribution is designing the temporal distribution of the prototype, we design two randomized groups: one to present MSTPNet's interpretation, and the other to present the interpretation without considering the temporal distribution which is in line with the state-of-the-art prototype learning methods. The only difference between the two groups is that MSTPNet considers the temporal distribution.

The first part of the user study collects seven control variables: age, education, gender, computer literacy, deep learning literacy, trust in AI, and medical literacy (Osborne et al., 2013). This user study passed randomization checks. The summary statistics of the final participants and randomization *p*-values are reported in Tables 16, 17, and 18.

**Table 16: Summary Statistics (Categorical)**

| Variable | Category | Count | Variable | Category | Count |
| --- | --- | --- | --- | --- | --- |
| Age | 18 and lower | 1 | Education | College freshman | 9 |
| | 18 – 24 | 21 | | College junior | 1 |
| | 25 – 34 | 32 | | College senior | 3 |
| | 35 - 44 | 1 | | Master | 21 |
| Gender | Female | 29 | | Doctorate | 21 |
| | Male | 26 | | | |

| Table 17: Summary Statistics (Continuous) | | | | | | |
|---|---|---|---|---|---|---|
| **Variable** | **Min** | **1st Qu.** | **Median** | **Mean** | **3rd Qu.** | **Max** |
| Computer Literacy | 1.000 | 2.000 | 3.000 | 2.727 | 3.000 | 4.000 |
| Deep Learning Literacy | 0.000 | 2.000 | 2.000 | 2.236 | 3.000 | 4.000 |
| Trust in AI | 1.000 | 2.000 | 3.000 | 2.527 | 3.000 | 4.000 |
| Health Literacy | 1.250 | 2.750 | 3.000 | 3.059 | 3.250 | 4.000 |

| Table 18: Randomization Checks | | | | | | | |
|---|---|---|---|---|---|---|---|
| | **Age** | **Education** | **Gender** | **Computer Literacy** | **Deep Learning Literacy** | **Trust in AI** | **Medical Literacy** |
| P-value | 0.433 | 0.534 | 0.921 | 0.488 | 0.788 | 0.944 | 0.894 |

Since the interpretation is based on depression symptoms and their temporal distribution, we design a depression knowledge education session for all participants. We ask the participants to read the background knowledge about depression diagnostic criteria. After reading such information, they are asked to answer questions. If they answer these questions incorrectly, an error message will prompt and direct them to read the information and choose again. After they answer questions correctly, they have sufficient knowledge to understand the interpretations.

| Table 19: The Interpretation of the Baseline Model | |
|---|---|
| **What Symptoms** | Evidence Posts |
| Suicidal thoughts | The first year of high school was insulted to the point of doubting life and wanting to commit suicide |
| Disturbed sleep | I didn't sleep well last night and almost took a whole bottle of sleeping pills last night |

| Table 20: The Interpretation of Our MSTPNet | | | | |
|---|---|---|---|---|
| **What symptoms** | **How Long** | | | |
| | **Frequency** | Evidence Posts | **Persistence** | Evidence Posts |
| Suicidal thoughts | 0.5 (times each week) | Use secret self-harm to relieve your inner pain. (04/13) | 2 (weeks) | Use secret self-harm to relieve your inner pain. (04/13) |
| | | The first year of high school was insulted to the point of doubting life and wanting to commit suicide. (04/21) | | The first year of high school was insulted to the point of doubting life and wanting to commit suicide. (04/21) |
| | | I'm leaving tonight, I hope there's no pain in the other world. (06/08) | | |
| Disturbed sleep | 0.3 (times each week) | I didn't sleep well last night and almost took a whole bottle of sleeping pills last night (05/02) | 1 (weeks) | I didn't sleep well last night and almost took a whole bottle of sleeping pills last night. (05/02) |
| | | I slept soundly. I woke up at 12 o'clock at noon. (01/08) | | |

Note: The red color is only used in the paper for demonstration purposes. The user study uses black color.

Then, we show the corresponding interpretation to each group separately, as shown in Tables 19 and 20. Subsequently, we ask the participants to rate their trust and perceived helpfulness of the given model, which are the common interpretability measurements for ML models (Xie et al., 2023). The measurement scale of trust is adopted from Chai et al. (2011), and Cronbach's Alpha is 0.901 for this scale, suggesting excellent reliability. The measurement scale of helpfulness is adopted from Adams et al. (1992). These scales also add an attention check question ("Please just select neither agree nor disagree"). After removing the participants who failed the attention check or the manipulation check, 55 participants remain in the final analyses. Table 21 shows the mean of trust and perceived helpfulness for each of the two groups. The participants' trust in our model (mean = 2.556) is significantly higher than the baseline model (mean = 1.659, $p < 0.001$). The participants' perceived helpfulness in our model (mean = 2.913) is significantly higher than the baseline model (mean = 2.567, $p = 0.023$).

**Table 21: Interpretability Comparison Between MSPTNet and Baseline**

| Interpretability measurement | MSTPNet (mean) | Baseline (mean) | MSTPNet (std) | Baseline (std) | P-value |
|---|---|---|---|---|---|
| Trust | 2.556 | 1.659 | 0.749 | 0.677 | < 0.001 |
| Perceived Helpfulness | 2.913 | 2.567 | 0.677 | 0.641 | 0.023 |

After the participants rate the trust and perceived helpfulness for the given model, we then show them the interpretation of the other model as a comparison. We ask them to choose a model interpretation that they trust more. 45 participants (81.8%) chose MSTPNet over the baseline. The above user study results prove that by interpreting the symptoms and their temporal distribution, MSTPNet improves users' trust and perceived helpfulness in our model, which offers empirical evidence for the contribution of our model innovation.

## DISCUSSION AND CONCLUSION

Depression detection is key for early intervention to mitigate economic and societal ramifications. The potential of social media data has attracted many scholars to develop a depression detection model. While deep learning-based models achieve superior performance, the lack of interpretability limits their uptake in high-stake decision-making scenarios such as healthcare. The interpretation provided by current interpretable deep learning methods mostly departs from the clinical depression diagnosis criterion that is based on depressive symptoms. Although a few studies have attempted to provide symptom-based interpretation, they face two limitations, including limited symptoms and neglecting the temporal distribution of symptoms.

Benefiting from the recent prototype learning, a class of state-of-the-art interpretable methods, we propose a novel interpretable deep learning method, MSTPNet, to detect and interpret depression based on what symptoms the user suffers and how long these related symptoms last. We propose a period-level analysis different from post-level and user-level and facilitate the analysis using a novel temporal segmentation layer, which segments user-generated posts into a period by considering both semantic similarity and the time interval between posts. Our proposed multi-scale temporal prototype layer can explicitly separate the task of identifying complex dynamic prototypes into two relatively simple tasks, which can effectively capture critical measurements (e.g., frequency, persistence) from the temporal distribution of depressive symptoms. We select a large-scale public dataset WU3D as our testbed. We conduct extensive evaluations to demonstrate the superior predictive power of our method over state-of-the-art benchmarks and showcase its interpretation of detection. Furthermore, through a user study, we show that our method outperforms these benchmarks in terms of interpretability.

**Contributions to the IS Knowledge Base**

Our study makes several contributions. First, our work belongs to the computational genre of design science research (Rai, 2017), which develops computational methods to solve business and societal problems and aims to make methodological contributions. In this regard, our proposed MSTPNet is a novel prototype learning method that processes a sequence of user-generated posts and interprets final decisions based on the type and temporal distribution of learned prototypes. To design MSTPNet, we innovatively overcome two main methodological challenges: the definition and segmentation of periods and the comprehensive measurements of the distribution of prototypes over periods. Second, our study contributes to healthcare IS research with a novel interpretable depression detection method, which detects depression and interprets its decision by showing what symptoms the user suffers and how long these symptoms last, which is aligned with the clinical depression diagnosis criterion. This method can also be generalized to other social media-based user-level detection, such as patients with anxiety disorders and fake reviewers. Third, our study contributes to medical research with new symptoms such as sharing negative events, which is discovered in data-driven methods.

Other than the instantiation of the MSTPNet approach in healthcare, our study also motivates, examines, and establishes five generalized design principles for design science research: (1) A temporal segmentation module could facilitate period-level analysis and mitigate the effect of redundant and irrelevant information; (2) Leveraging prototype learning method could discover new knowledge (e.g., depressive symptoms) from user-generated content; (3) It's cost-effective and flexible to explicitly separate a complex task into two related simple tasks, e.g., identifying simple prototypes and analyzing these prototypes according to a specific objective; (4) Combining information on multiple time scales could comprehensively understand the predicted subject and compensates the vulnerabilities at a single scale; (5) Showing the temporal distribution of prototypes could improve interpretability, and boost the trust and perceived helpfulness. These design principles prescribe how to predict and interpret the hidden state of a user from a sequence of user-related data, such as social media posts, electric health records, and sensor-based signals. The prescriptive knowledge is generalizable to other predictive analytics contexts.

## Practical Implications

Our proposed method holds significant practical implications. For government agencies, our method can estimate the number of people with depression in different regions and shows the prevalence of each depressive symptom. This analysis can support the design of public health policies. Compared with a questionnaire, social media-based depression detection is faster and cheaper. Our method can be deployed in social media platform to personally recommend online resources such as articles and videos according to specific symptoms of target users. For nongovernmental organizations or volunteers, our method can help them find depressed patients to provide information and emotional support by sending replies or messages.

## Limitations and Future Research

Our study has a few limitations and can be extended in future work. First, users' posts are often irregular and difficult to cover all the time. Therefore, after segmenting, two consecutive periods may not be continuous in reality, which may affect the calculation of the persistence of the symptom prototype. This is a shared issue with most other deep learning methods such as RNN and CNN. Future work could impute the missing values in the time dimension or consider the time distance between different periods. Second, correlations between different depression symptoms are not considered in our proposed method. In the actual diagnosis of depression, there may be a hidden link between symptoms, and depression and other chronic diseases often appear together. Future research can build upon our model and consider the impact of co-occurrence between symptom on the model's predictive ability without loss of interpretability.